\documentclass[authoryear,12pt]{article}

\usepackage{graphicx} 
\usepackage{amsmath}
\usepackage{natbib} 
\usepackage{url}
\usepackage{epstopdf}
\usepackage{multirow}
\usepackage[]{algorithm, algorithmic}
\usepackage{ulem} 
\usepackage{color} 

\providecommand{\keywords}[1]
{
	\small	
	\textbf{\textit{Keywords---}} #1
}

\newcommand{\argmax}{\mathop{\mathrm{argmax}}\limits}
\newcommand{\argmin}{\mathop{\mathrm{argmin}}\limits}
\newcommand{\minF}{\mathop{\mathrm{min}}\limits}
\newcommand{\maxF}{\mathop{\mathrm{max}}\limits}
\newcommand{\aveF}{\mathop{\mathrm{ave}}\limits}

\newcommand{\R}[1]{\texttt{#1}}

\begin{document}

	\title{Musings on Constructions of Optimal Latin Hypercube Designs with Flexible Sizes}
	
	\author{Hongzhi Wang, Qian Xiao, Abhyuday Mandal}
	
	\date{}
	\maketitle
	
	\begin{abstract}

        Latin hypercube designs (LHDs) play an important role in computer experiments, offering flexible and efficient space-filling properties under a variety of optimality criteria, including maximin distance, maximum projection, and orthogonality. Constructing optimal LHDs with flexible sizes is challenging due to the limited availability of theoretical results, namely, algebraic constructions with provable optimality guarantees, and the rapidly expanding search space encountered by search algorithms. For design sizes outside the scope of known algebraic constructions, search-based algorithms are widely used, but they often involve substantial computational effort and lack assurances of global optimality. This paper provides a comprehensive review and comparison of current popular algebraic constructions for optimal LHDs and widely adopted search algorithms for generating high-quality LHDs. By reviewing their theoretical properties and empirical performance across a range of criteria, we offer a unified perspective on their relative strengths and limitations. The comparisons presented herein aim to assist practitioners in selecting appropriate design strategies for different objectives and constraints. The insights from this work also highlight open challenges and may serve as a benchmark for future development in optimal LHD construction.
			
	\end{abstract}

	\keywords{Computer Experiments; Space-filling design; Maximum projection design; Orthogonality; Simulated Annealing; Particle Swarm Optimization; Genetic Algorithm.}

	\section{Introduction}
	
	Computer experiments are widely used in scientific research and industrial applications, where complex computer codes, often high-fidelity simulators, are used to generate data instead of physical experimentation \citep{sacks1989designs, fang2005design}. Because their outputs are deterministic and free of random errors, replication is unnecessary \citep{butler2001optimal, joseph2008orthogonal, ba2015optimal}. Latin hypercube designs (LHDs,  \cite{mckay1979comparison}), characterized by their avoidance of replication in each dimension and their uniform one-dimensional projections, may be regarded the most popular design class for computer experiments \citep{fang2005design, xiao2018construction}. Depending on practical objectives, different forms of optimal LHDs are used, including space-filling, maximum projection, and orthogonal designs. Although numerous construction methods have been developed, obtaining high-quality LHDs for moderate to large design sizes remains challenging \citep{ye1998orthogonal, fang2005design, joseph2015maximum, xiao2018construction}. One challenge is that theoretical results, such as algebraic constructions guaranteeing properties like maximin distance or orthogonality, only exist for certain design sizes, which limits their general applicability. For sizes where such guarantees are unavailable, search algorithms are typically employed; however, their performance varies with the choice of algorithm, the extent of the search space explored, and the computational resources allocated, meaning global optimality cannot be ensured. More fundamentally, optimal LHD construction is a discrete optimization problem, and although exhaustive enumeration would identify the true optimum, it becomes infeasible for all design sizes due to the exponential growth of the permutation space.
	
	An LHD with $n$ runs and $k$ factors is represented by an $n \times k$ matrix in which each column is a permutation of integers $1, \ldots, n$. Throughout this paper, $n$ refers to the run size and $k$ refers to the factor size. Space-filling LHDs aim to distribute design points as uniformly as possible across the input domain, thereby reducing unsampled regions in all dimensions. Various criteria have been proposed to quantify this property, including maximin and minimax distance measures \citep{johnson1990minimax, morris1995exploratory}, different forms of discrepancy \citep{hickernell1998generalized, fang2002centered, fang2005design}, and entropy-based measures \citep{shewry1987maximum}. Because the number of possible LHDs increases exponentially as $(n!)^{k}$ for a given $n$ and $k$, an exhaustive enumeration is impractical for all design sizes. The current literature contains search algorithms \citep{morris1995exploratory, leary2003optimal, joseph2008orthogonal, ba2015optimal, kenny2000algorithmic, jin2005efficient, liefvendahl2006study, grosso2009finding, chen2013optimizing} and algebraic constructions \citep{zhou2015space, xiao2017construction, wang2018optimal} designed to produce space-filling LHDs.

    Search algorithms are commonly used to construct space-filling LHDs with flexible run sizes. For example, \cite{morris1995exploratory} proposed a simulated annealing (SA) algorithm that helps avoid being trapped in local optima and improves the likelihood of finding globally optimal designs. Building on their work and that of \cite{tang1993orthogonal}, \cite{leary2003optimal} developed an SA-based approach for searching space-filling orthogonal array–based LHDs (OALHDs). \cite{joseph2008orthogonal} introduced a multi-objective criterion and an adapted SA algorithm that simultaneously accounts for orthogonality and space-filling properties. \cite{ba2015optimal} extended the framework of sliced Latin hypercube designs (SLHDs; \cite{qian2012sliced}) and proposed a two-stage SA algorithm.
    Beyond SA-based algorithms, several other search algorithms have been developed to identify optimal designs. \cite{kenny2000algorithmic} proposed the columnwise pairwise (CP) algorithm for generating efficient symmetric LHDs. \cite{jin2005efficient} introduced the enhanced stochastic evolutionary (ESE) algorithm, which combines an exchange procedure (inner loop) with a threshold-determination mechanism (outer loop). \cite{liefvendahl2006study} employed a genetic algorithm (GA) with a strategy that directly updates toward the global best. \cite{grosso2009finding} explored an iterated local search heuristic that alternates between local improvement and perturbation. \cite{chen2013optimizing} developed a particle swarm optimization (PSO) algorithm in which the search process gradually reduces the Hamming distance between each particle and the global or personal best through element exchanges. In general, search algorithms can be computationally intensive, especially for constructing large designs.
    For certain design sizes, algebraic constructions provide an attractive alternative, often yielding optimal LHDs at very low computational cost. For instance, \cite{xiao2017construction} used Costas arrays to construct space-filling saturated LHDs or Latin squares with run sizes $n=p-2$, $p-1$ or $p$, where $p$ is any prime or prime power. \cite{wang2018optimal} applied the Williams transformation \citep{williams1949experimental} to linearly permuted good lattice point sets to obtain maximin LHDs when the number of factors does not exceed the count of positive integers that are coprime to the run size.

    While traditional space-filling designs emphasize coverage of the full dimensional input space, it is often equally important to ensure space-filling properties in intermediate dimensions. To address this need, \cite{joseph2015maximum} introduced maximum projection (MaxPro) designs, which enhance space-filling performance across all sub-spaces of dimension 2 through $k-1$. Compared with classical maximin distance LHDs, MaxPro LHDs typically yield better separation in these lower-dimensional views. However, generating high-quality MaxPro LHDs remains challenging, particularly for large design sizes. \cite{joseph2015maximum} proposed an SA-based algorithm for their construction. In this work, we explore a GA framework that is capable of producing MaxPro LHDs with substantially improved performance, and illustrative examples are provided in Section~\ref{Result}.

    Unlike space-filling LHDs, which aim to reduce similarity among design points (rows), orthogonal LHDs (OLHDs) focus on minimizing dependence among factors (columns). OLHDs are particularly appealing because they achieve zero column-wise correlation, thereby facilitating more reliable estimation of factor effects. Algebraic constructions have been developed for specific design sizes. For example, \cite{ye1998orthogonal} proposed an algebraic construction of OLHDs with run sizes $n=2^m+1$ and factor sizes $k=2m-2$, where $m$ is any integer no less than 2. \cite{cioppa2007efficient} further extended this approach to accommodate more factors. \cite{steinberg2006construction} developed a construction method based on factorial designs and group rotations for designs with $n=2^{2^m}$ and $k=2^{m}t$, where $m$ is any positive integer and $t$ is the number of rotation groups. \cite{sun2010construction} extended their earlier work \citep{sun2009construction} to construct OLHDs with $n=r2^{c+1}$ or $n=r2^{c+1}+1$ and $k=2^c$, where $c$ and $r$ are positive integers. \cite{yang2012construction} proposed using generalized orthogonal designs to construct OLHDs and nearly orthogonal LHDs (NOLHDs) with $n=2^{r+1}$ or $n=2^{r+1}+1$ and $k=2^r$, where $r$ is any positive integer. \cite{georgiou2014some} introduced a construction method based on orthogonal arrays and their full fold-overs, yielding designs with $n=2ak$ runs and $k$ factors, where $k$ is the dimension of the orthogonal matrix and $a$ is any positive integer. \cite{butler2001optimal} applied the Williams transformation \citep{williams1949experimental} to construct OLHDs under a second-order cosine model for odd prime $n$ and $k\leq n -1$.
    In addition, several new types of optimal LHDs can be derived by leveraging properties of orthogonal arrays. For example, \cite{tang1993orthogonal} proposed to construct OALHDs by deterministically replacing elements in orthogonal arrays, resulting in designs with improved space-filling properties and low column-wise correlations. \cite{lin2009construction} proposed using OLHDs or NOLHDs as starting designs and coupling them with orthogonal arrays to obtain OALHDs. Their approach requires fewer runs to support large numbers of factors, producing designs with $n^2$ runs and $2fp$ factors, where $n$ and $p$ are design sizes of the OLHDs or NOLHDs, and $2f$ is the number of columns in the coupled orthogonal array.
    However, many design sizes fall outside the scope of these theoretical results, and in such cases OLHDs can be obtained through search algorithms. The optimization algorithms discussed above can be adapted to the target orthogonality.
    
    In this paper, we provide a comprehensive review and comparison of current popular algebraic constructions for optimal LHDs and widely adopted search algorithms for generating high-quality LHDs. By examining their theoretical properties and empirical performance across a variety of criteria, we present a unified perspective on their respective strengths and limitations. The comparisons offered here are intended to guide practitioners in selecting design strategies suited to different objectives and constraints. In conducting these evaluations, we also identify several designs that outperform existing ones; these improved designs are reported in Section~\ref{Result}.

    The remainder of the paper is organized as follows. Section~\ref{OC} introduces different optimality criteria for LHDs. Section~\ref{Algs} describes some widely used search algorithms and their implementations. Section~\ref{Constr} presents several popular algebraic constructions. Section~\ref{Result} reports numerical results and comparisons. Section~\ref{Con} concludes with a discussion and potential directions for future research.
	
	\section{Optimality Criteria for LHDs}\label{OC}

    A variety of criteria have been proposed to quantify designs' space-filling performances \citep{johnson1990minimax, hickernell1998generalized, fang2002centered, fang2005design, shewry1987maximum}. Among these, the maximin distance criterion \citep{johnson1990minimax} has become one of the most widely used. It aims to distribute design points throughout the experimental region so that the smallest pairwise distance between points is as large as possible, thereby promoting uniform coverage. Let $\textbf{X}$ denote an LHD matrix throughout this paper. For two runs $x_i$ and $x_j$ of $\textbf{X}$, the $L_q$-distance is defined as $d_q(x_i, x_j) =  \left\{ \sum_{m=1}^{k} \vert x_{im}-x_{jm}\vert ^q \right\}^{1/q}$, where $q$ is an integer, $1 \leq i \leq n$, $1 \leq j \leq n$, and $i \neq j$. In this paper, we focus on the cases $q=1$ and $q=2$, corresponding to the Manhattan ($L_1$) and Euclidean ($L_2$) distances. The overall $L_q$-distance of the design $\textbf{X}$ is defined as $d_q(\textbf{X}) = \text{min} \{d_q(x_i, x_j),  1 \leq i<j \leq n \}$. A design $\textbf{X}$ is said to be a maximin $L_q$-distance design if it achieves the largest $d_q(\textbf{X})$ among all LHDs of the same size. If multiple designs share the same largest $d_q(\textbf{X})$, the maximin distance design sequentially maximizes the next minimum inter-site distances. To facilitate numerical optimization of the maximin distance criterion, \citet{morris1995exploratory} and \citet{jin2005efficient} proposed to minimize a scalar value:
	\begin{equation}\label{E1}
		\phi_{p}= \bigg\{\sum_{i=1}^{n-1}\sum_{j=i+1}^{n}d_q(x_i, x_j)^{-p} \bigg\} ^{1/p},
	\end{equation}
	where $p$ is a tuning parameter. This $\phi_{p}$ criterion in Equation~\eqref{E1} approaches the maximin distance criterion as $p \to \infty$. In practical applications, choosing $p=15$ typically suffices \citep{morris1995exploratory}.
    
	Maximin distance LHDs primarily emphasize the space-filling properties in the full $k$-dimensional design space, but they do not necessarily ensure good separation when designs are projected onto lower-dimensional subspaces. To address this limitation, \cite{joseph2015maximum} introduced the maximum projection (MaxPro) criterion, which explicitly evaluates space-filling performance in all possible projections. An LHD $\textbf{X}$ is a MaxPro LHD if it minimizes  
    \begin{equation}\label{E2}
        \minF_{\textbf{X}} \psi (\textbf{X}) = \Bigg\{ \frac{1}{\binom{n}{2}} \sum_{i=1}^{n-1} \sum_{j=i+1}^{n} \frac{1}{\prod_{l=1}^{k}(x_{il}-x_{jl})^2}  \Bigg\}^{1/k}.
    \end{equation}
    Equation~\eqref{E2} penalizes pairs of design points that are close in any coordinate, thereby encouraging separation in every marginal projection. Therefore, MaxPro LHDs promote uniformity not only in the full-dimensional space but also across all lower-dimensional subspaces. It should be noted that the original MaxPro criterion was defined for designs scaled to the unit hypercube $[0,1]^{k}$ \citep{joseph2015maximum}, whereas the designs considered here use integer-valued levels. The two representations are equivalent up to a simple linear transformation. For instance, scaling can be reversed by applying the transformation $\textbf{X}_{Integer} = \textbf{X}_{Scaled}*n-0.5$, which maps unit-hypercube points to integer-level designs. This transformation is implemented in Section~\ref{Result} to ensure a fair comparison of the design performance.

    Orthogonal and nearly orthogonal designs, which seek to reduce the dependence among factors, are widely employed in experiments \citep{georgiou2009orthogonal, steinberg2006construction, sun2017general}. The degree of orthogonality in a design is commonly quantified using correlation-based measures. Two frequently used criteria are the average absolute correlation and the maximum absolute correlation \citep{georgiou2009orthogonal}, denoted by ave$(|q|)$ and max$|q|$, respectively:
    \begin{equation}\label{E3}
        \aveF(|q|) = \frac{2 \sum_{i=1}^{k-1} \sum_{j=i+1}^{k}|q_{ij}|}{k(k-1)} \text{ and } \maxF|q| =  \maxF_{i,j} |q_{ij}|,
    \end{equation}
    where $q_{ij}$ represents the correlation between the $i$th and $j$th columns of the design matrix $\textbf{X}$. A design is considered orthogonal if both ave$(|q|)$ and max$|q|$ equal zero, a condition that cannot be achieved for many combinations of run sizes and factor dimensions. In practical applications, designs with small values of either criterion are generally favored.

	\section{Search Algorithms for Constructing LHDs with Flexible Sizes}\label{Algs}
	
	\subsection{Simulated Annealing Based Algorithms} 
	
	Simulated annealing (SA, \cite{kirkpatrick1983optimization}) is a stochastic optimization technique inspired by the physical annealing process in metallurgy, in which a material is heated to a high temperature and then slowly cooled to reduce structural defects. As a single solution based algorithm, SA begins from an initial candidate solution and iteratively explores the search space, accepting or rejecting new solutions according to a probabilistic rule until convergence is reached. The algorithm has been widely applied to optimization problems involving continuous variables. To accommodate the discrete nature of LHDs, the classical SA framework has been adapted by researchers to enable effective search over the permutation-based design space. In this article, we focus on minimizing the optimality criteria outlined in Section~\ref{OC}, meaning only minimization optimization problems are considered.

    \citet{morris1995exploratory} proposed a modified simulated annealing algorithm in which elements of an LHD are randomly exchanged to search for potential improvements. Specifically, the algorithm begins with a randomly generated LHD, from which a column is selected at random. Then, two random elements within this column are swapped to produce a new candidate design. For example, consider an LHD with four runs in which a selected column contains the elements $(1,2,3,4)$. Exchanging the first two elements yields the column  $(2,1,3,4)$, resulting in a new LHD. If this exchange improves the design under a specified optimality criterion, it is accepted; otherwise, it is accepted with probability $\hbox{exp}[-(\Phi(\textbf{X}_{new})-\Phi(\textbf{X}))/T]$, where $\Phi$ denotes the optimality criterion, $\textbf{X}$ is the current design,  $\textbf{X}_{new}$ is the design obtained after the exchange, and $T$ is the current temperature. This exchange procedure is applied iteratively to progressively improve the design. When no improvement is observed after a prescribed number of attempts, the temperature $T$ is reduced according to the annealing schedule, thus decreasing the probability of accepting inferior designs and promoting convergence. The algorithm ends when convergence is achieved or a predefined computational budget is exhausted, at which point the best LHD encountered is returned. The pseudo-code of this SA is outlined in Algorithm~\ref{Alg1}.
	
	\begin{algorithm}
		\caption{Simulated Annealing Algorithm for LHDs}
		\begin{algorithmic}[1]\label{Alg1}
			\STATE Generate a random LHD, denoted as $\textbf{X}$.
			\WHILE {not converge}
			
			\STATE Randomly choose a column from $\textbf{X}$, denoted as $j$.
			\STATE Exchange two randomly selected elements within column $j$, and denote the new LHD as $\textbf{X}_{new}$.
			\STATE If $\Phi(\textbf{X}_{new}) < \Phi(\textbf{X})$, then $\textbf{X}=\textbf{X}_{new}$. Otherwise, let $\textbf{X}=\textbf{X}_{new}$ with a probability of $\hbox{exp}[-\frac{\Phi(\textbf{X}_{new})-\Phi(\textbf{X})}{T}]$.
			\STATE When no improvements are found consecutively for certain attempts, decrease the current temperature $T$ and repeat Steps 3$-$5.
			
			\ENDWHILE
		\end{algorithmic}
	\end{algorithm}

    \citet{leary2003optimal} extended the simulated annealing algorithm of \citet{morris1995exploratory} to construct orthogonal array–based Latin hypercube designs (OALHDs). \citet{tang1993orthogonal} demonstrated that OALHDs generally exhibit superior space-filling properties compared to randomly generated LHDs. The algorithm proposed by \citet{leary2003optimal} begins with a randomly selected OALHD, from which a column is chosen at random. Then, two elements within this column that correspond to the same entry in the underlying orthogonal array (OA) are exchanged. For instance, consider an OALHD with nine runs in which a selected column contains elements $(1,2,3,4,5,6,7,8,9)$, and the corresponding OA has entries $({\it 1,1,1},{\it 2,2,2},{\it 3,3,3})$. In this case, elements $(1,2,3)$ share the same original OA entry ${\it 1}$, elements $(4,5,6)$ share the entry ${\it 2}$, and elements $(7,8,9)$ share the entry ${\it 3}$. The remaining steps of the algorithm follow those of the simulated annealing procedure described in \cite{morris1995exploratory}. It is important to note that the existence of OALHDs depends on the availability of the corresponding underlying orthogonal arrays.

    \citet{joseph2008orthogonal} proposed another variant of simulated annealing for constructing orthogonal–maximin LHDs, which simultaneously account for orthogonality and maximin distance criteria. The algorithm begins by generating a random LHD and then identifies the column exhibiting the largest average pairwise correlation with all other columns. Specifically, the $l^{*}$th column is selected, where $l^{*} = \argmax \rho_{l}^2$ with $\rho_{l}^2=\frac{1}{k-1}\sum_{j\neq l}\rho_{lj}^2$, and $\rho_{lj}$ denotes the correlation between the $l$th and $j$th columns. Next, the algorithm selects the row with the largest total row-wise distance from all other rows. That is, the  $i^{*}$th row is chosen such that $i^{*}=\argmax (\sum_{j\neq i}d(x_i, x_j)^{-p})^{1/p}$, where $d(x_i, x_j)$ represents the distance between the $i$th and $j$th rows. The element located at the intersection of the $i^{*}$th row and the selected column is then swapped with a randomly chosen element of the same column. The remaining steps of the procedure follow the simulated annealing framework described in \cite{morris1995exploratory}. Although this algorithm is capable of producing LHDs that simultaneously minimize the $\phi_{p}$ and ave$(|q|)$ criteria, it is computationally intensive, as all $\rho_{l}^2$ values and row-wise distances must be recomputed at each iteration. Table~\ref{T0} summarizes the similarities and differences among the three simulated annealing algorithms discussed above.
	
	\begin{table}
		\caption{Summary Table of Simulated Annealing Based Algorithms.}\label{T0}
		\begin{center}
			\resizebox{\textwidth}{!}{\begin{tabular}{|l|c|c|c|}
					\hline
					&SA \citep{morris1995exploratory}&OASA \citep{leary2003optimal}&SA2008 \citep{joseph2008orthogonal}\\
					
					\hline
					\multirow{3}{*}{Starting Design}
					&&&\\
					&A random LHD&A random OALHD&A random LHD \\
					&&&\\
					
					\hline
					\multirow{3}{*}{Column Choice}
					&&&A column who has\\
					&A random column&A random column& the largest average\\
					&               &               & pairwise correlation\\
					
					\hline
					\multirow{3}{*}{Elements Choice}
					&&Two random elements&The element from the row having\\
					&Two random elements&share the same&largest total row-wise distance with a \\
					& &original OA entry &random element in the same column\\
					
					\hline
			\end{tabular}}
		\end{center}
	\end{table}

    \cite{ba2015optimal} and \cite{qian2012sliced} proposed methods for constructing space-filling sliced Latin hypercube designs (SLHDs). An SLHD with $n$ runs can be partitioned into $t$ slices, $\textbf{X}=\bigcup_{i=1}^{t}\textbf{X}_{i}$, where each slice $\textbf{X}_{i}$ is itself an LHD with $m=n/t$ runs. They introduced a simulated annealing algorithm consisting of two stages. The first stage optimizes the design from the slice perspective and begins with $t$ randomly generated slices, each being an $m \times k$ LHD. If duplicated rows are detected across slices, the algorithm randomly selects one duplicated row, then chooses another row within the same slice, and swaps the two elements in these rows for a randomly selected column. This procedure is repeated until no duplicate rows remain. Subsequently, elements in a randomly chosen column of a randomly selected slice are exchanged. If this exchange results in duplicated rows, the previous procedure is invoked again; otherwise, the exchange is retained when it leads to an improvement. The second stage refines the design at the element level by replacing the $t$ occurrences of the level $l$ ($l=1, \ldots, m$) with a random permutation of $\{ (l-1)t+1,\ldots, lt \}$, followed by swapping two randomly selected elements within that permutation. When $t=1$, the algorithm reduces to a procedure for constructing classical space-filling LHDs.

	\subsection{Particle Swarm Optimization Algorithms}\label{LaPSo}

    Particle swarm optimization (PSO, \cite{kennedy1995particle}) is a metaheuristic optimization algorithm inspired by collective behaviors observed in nature, such as coordinated flight patterns of bird flocks. PSO is a population-based algorithm in which a set of particles, representing candidate solutions, explores the search space by updating their positions and velocities according to predefined mathematical rules, with the goal of iteratively improving a specified optimality criterion. The movement of each particle is influenced by both its own best-known position (personal best) and the best positions discovered by the swarm as a whole (global best). Through this cooperative information-sharing mechanism, the swarm is expected to converge toward high-quality solutions.
    
    Recent studies \citep{chen2013optimizing, chen2015minimax, wong2015modified} have adapted the classical PSO framework to the construction of high‑quality experimental designs, a problem characterized by a discrete search space. In particular, \citet{chen2013optimizing} proposed a PSO-based algorithm, known as LaPSO, to identify maximin distance LHDs. This approach modifies the standard PSO update rules for particle velocities and positions. Specifically, it seeks to reduce the Hamming distance between each particle and the global best (or its corresponding personal best) by deterministically exchanging elements. Here, the Hamming distance between two LHD particles is defined as the number of positions at which their corresponding elements differ.
	
	LaPSO begins by generating $m$ random LHDs, where $m$ denotes the number of particles and serves as a tuning parameter. Initially, the personal best (PB) of each particle is set to the particle itself, while the global best (GB) is defined as the particle LHD that achieves the best value under a specified optimality criterion $\Phi$. The algorithm then applies an exchange procedure to each column of every particle LHD. Specifically, for a given column, an element in the current particle (denoted as $e_r$)is randomly selected, and the corresponding element located at the same position in the PB (denoted as $e_p$) is identified. The elements $e_r$ and $e_p$ are then swapped within the current column, thereby reducing the Hamming distance between the current particle and its PB. An analogous exchange procedure is performed to reduce the Hamming distance between the current particle and the GB. These two operations emulate particle movement toward the PB and the GB, respectively.

    Either of these steps may be skipped or repeated, but they cannot be omitted simultaneously. Although no strict upper bound is imposed on the number of repetitions, performing a large number of exchanges, i.e. $2n$, will make the current column nearly identical to that of the PB or the GB. After all particle LHDs have been updated, a random number $z$ is drawn from the standard uniform distribution for each column of each particle. If $z < p_0$, where $p_0$ is a user-defined tuning parameter, two randomly selected elements in the current column are exchanged. This random perturbation helps prevent premature convergence to local optima. Once the algorithm converges or the computational budget is exhausted, the GB is returned as the final solution. Unlike the exchange mechanisms used in SA based algorithms, LaPSO explicitly links each particle to its personal best, the global best, or both. However, since each exchange operation can reduce the Hamming distance by at most one, LaPSO may become computationally demanding when constructing large high‑quality designs. The pseudo-code of LaPSO is outlined in Algorithm~\ref{Alg2}.
	
	\begin{algorithm}
		\caption{LaPSO}
		\begin{algorithmic}[1]\label{Alg2}
			\STATE Generate $m$ random LHDs, denoted as $L_{1},\ldots, L_{m}$.
			\STATE Initialize the personal best. Set $PB_{i}=L_{i}$ for $i=1,\ldots,m$.
			\STATE Initialize the global best. Set $GB=\argmin_{i} \Phi(L_{i})$.
			
			\WHILE {not converge}
			
			\FOR{each particle LHD $L_{i}$}
			\FOR{each column $j$ of $L_{i}$}
			\STATE Randomly choose an element in column $j$, denoted as $e_r$, and then find the element in $PB_{i}$ whose location is the same as $e_r$, denoted by $e_p$. Swap $e_r$ and $e_p$ within the column $j$.
			
			\STATE Randomly choose an element in column $j$, denoted as $e_r$, and then find the element in $GB$ whose location is the same as $e_r$, denoted by $e_g$. Swap $e_r$ and $e_g$ within the column $j$.
			
			\IF{ $z<p_0$, where $z \sim$ Uniform(0,1)}
			\STATE Exchange two randomly selected elements in column $j$.
			\ENDIF
			
			\ENDFOR
			\ENDFOR
			
			\FOR {each updated $L_{i}$}
			
			\IF {$\Phi(L_{i}) < \Phi(PB_{i})$}
			\STATE $PB_{i}=L_{i}$.
			\ENDIF
			
			\IF {$\Phi(L_{i}) < \Phi(GB)$}
			\STATE $GB=L_{i}$.
			\ENDIF
			
			\ENDFOR
			
			\ENDWHILE
			
		\end{algorithmic}
	\end{algorithm}

	\subsection{Genetic Algorithms}\label{GA}

    Genetic algorithms (GAs) are nature-inspired metaheuristic optimization algorithms that emulate the principle of natural selection proposed by Charles Darwin \citep{holland1992adaptation, goldberg1989genetic}. A GA is a population-based algorithm that typically consists of stages of ${\it selection}$, ${\it crossover}$, ${\it mutation}$ and ${\it fitness}$ ${\it evaluation}$. In accordance with evolutionary terminology, candidate solutions are referred to as chromosomes, which together constitute a population. Based on their performance under a specified criterion, called the fitness, a subset of chromosomes is selected as parents for generating a new population, a process known as selection. The selected parents are then recombined through crossover and mutation operators to produce offspring that are expected to exhibit improved performance.
	
	\cite{liefvendahl2006study} proposed a GA for constructing maximin distance LHDs. The initial population consists of $m$ randomly generated LHDs, where $m$ is an even, user-defined population size. During the ${\it selection}$ stage, the entire population is evaluated according to the specified optimality criterion, and the top half of the population is retained as ${\it survivors}$. Among these, the current global best is designated as ${\it best}$ ${\it survivor}$ (BS). The BS and the remaining survivors are then used to generate a new population.
    
    Specifically, the BS is assigned as the first individual, and the second through the $(m/2)$th individuals are created via the following ${\it crossover}$ procedure. For each survivor other than the BS, a column is randomly selected from that survivor and used to replace the column with the same index in the BS, thereby generating a new individual. For the remaining half of the population, the BS is designated as the $(m/2+1)$th individual, and the $(m/2+2)$th through the $m$th individuals are generated through an alternative ${\it crossover}$ operation. In this case, for each survivor except the BS, a column is randomly selected from the survivor and replaced by the column with the same index in the BS to form a new individual.
    
    After constructing the new population of $m$ LHDs, a ${\it mutation}$ step is applied to reduce the risk of premature convergence. During ${\it mutation}$, for each column of every new LHD except the first design, a random number $z$ is drawn from the standard uniform distribution. If $z < p_{mut}$, where $p_{mut}$ is a user-defined tuning parameter, two randomly selected elements within that column are exchanged. Next, the entire population is then re-evaluated to identify the survivors for the next iteration. Once the algorithm converges or the computational budget is exhausted, the BS is returned as the final design. 

    Unlike LaPSO, which relies on element-wise exchanges, this GA operates by replacing entire columns in both the current global best design and candidate designs. Consequently, the Hamming distance for the replaced columns becomes zero immediately with respect to the global best. Compared with other search algorithms, this approach typically requires less CPU time, particularly for large design sizes. The pseudo-code of GA is outlined in Algorithm~\ref{Alg3}.

	\begin{algorithm}
		\caption{Genetic Algorithm for LHD}
		\begin{algorithmic}[1]\label{Alg3}
			\STATE Generate $m$ random LHDs, denoted as $L_{1},\ldots, L_{m}$, where $m$ is an even number.
			\STATE Calculate $\Phi(L_{i})$ for $i=1,\ldots, m$, and select the best $\frac{m}{2}$ $L_{i}$ (with the smallest $\frac{m}{2}$ $\Phi$ values), denoted by $L_{i}^{s}$ for $i=1,\ldots, \frac{m}{2}$, WLOG.
			\STATE Identify the ${\it best}$ ${\it survivor}$, $L_{b}^{s}$, i.e. $L_{b}^{s}=\argmin_{i} \Phi(L_{i}^{s})$.
			
			\WHILE {not converge}
			
			\STATE Let $L_{1}=L_{b}^{s}$ and $c=2$.
			\FOR{each $L_{i}^{s}$ (except $L_{b}^{s}$)}
			
			\STATE Randomly choose a column $j$ from $L_{i}^{s}$, and use it to replace the $j$th column of $L_{b}^{s}$. Let this new matrix be $L_{c}$, and $c=c+1$.
			
			\ENDFOR
			
			\STATE Let $L_{m/2+1}=L_{b}^{s}$ and $c=m/2+2$.
			\FOR{each $L_{i}^{s}$ (except $L_{b}^{s}$)}
			
			\STATE Randomly choose a column $j$ from $L_{i}^{s}$, and replace it with the $j^{th}$ column from $L_{b}^{s}$. Let this new matrix be $L_{c}$, and $c=c+1$.
			
			\ENDFOR
			
			\FOR{each $L_{i}$ (except $L_{1}$)}
			\FOR{each column $j$ of $L_{i}$}
			\IF{$z < p_{mut}$, where $z \sim \text{Uniform}(0,1)$ }
			\STATE Exchange two randomly selected elements in column $j$.
			\ENDIF
			\ENDFOR
			\ENDFOR
			
			\STATE  Repeat Steps $2-3$.
			
			\ENDWHILE
			
		\end{algorithmic}
	\end{algorithm}

	\section{Algebraic Constructions for Optimal LHDs with Certain Sizes}\label{Constr}

    For certain design sizes, algebraic constructions are available, and theoretical results have been developed to guarantee the efficiency of the resulting designs. Such constructions require little to no search and are therefore especially attractive for designs with large sizes. In this section, we review several practically useful algebraic constructions for maximin distance LHDs and orthogonal LHDs.
	
	\subsection{Algebraic Constructions for Maximin Distance LHDs}\label{WT}
	
	\cite{wang2018optimal} proposed to generate maximin distance LHDs via good lattice point (GLP) sets \citep{zhou2015space} and Williams transformation \citep{williams1949experimental}. In practice, their method can lead to space-filling designs with relatively flexible sizes, where the run size $n$ is flexible but the factor size $k$ must be no greater than the number of positive integers that are co-prime to $n$. They proved that the resulting designs of sizes $n \times (n-1)$  (with $n$ being any odd prime) and $n \times n$ (with $2n+1$ or $n+1$ being odd prime) are optimal under the maximin $L_1$ distance criterion. 
	
	The construction method in \cite{wang2018optimal} can be summarized into three steps. First, generate an $n \times k$ GLP design $D$ whose element is $x_{ij}=i \times h_{j}$ (mod $n$) with $i=1, \ldots, n$, $j=1, \ldots, k$ and $h=(h_1, \ldots,h_k)$ being a set of distinct positive integers that are coprime to $n$. Second, for any $b\in  \left\{0,\ldots,n-1\right\}$, generate $D_b=D+b \text{ (mod } n\text{)}$ and $E_b=W(D_b)$, where $W: \mathcal{Z}_n \rightarrow \mathcal{Z}_n$ is the Williams transformation \citep{williams1949experimental}:
	\begin{equation}
		\label{wt}
		W(x)=\left\{\begin{array}{ll}
			2x, & \ 0\leq x \leq (n-1)/2, \\
			2(n-x)-1, & \ (n+1)/2\leq x \leq n-1.
		\end{array}
		\right.
	\end{equation}
	Third, find the best $b^{*} \in \{ 0,\ldots,n-1\}$ such that $E_{b^{*}}$ has the smallest $\phi_p$ value.

    As shown in \cite{zhou2015space} and \cite{wang2018optimal}, the designs $D$ and $E$ defined above are $n \times k$ LHDs whose last rows contain the same elements. By removing the last rows and reordering the levels, the remaining designs can be transformed into LHDs, referred to as leave-one-out designs $\tilde{D}$ and $\tilde{E}$. After performing the three steps outlined above, the resulting designs $E_{b^{*}}$ and $\tilde{E}_{b^{*}}$ exhibit good space-filling properties. The construction method proposed by \cite{wang2018optimal} is particularly attractive for generating large maximin distance LHDs; however, it may be too complex for practitioners to implement directly. In the \R{LHD} package in R, the function \R{FastMmLHD()} implements this method with a user-friendly interface.
	
	\cite{tang1993orthogonal} proposed to construct orthogonal array-based LHDs (OALHDs) from existing orthogonal arrays (OAs). Suppose  $\textbf{A}$=OA($n,m,s,r$) is an orthogonal array with $n$ rows, $m$ columns, $s$ levels ($n > s \geq 2$) and $r$ strength. If each $n \times r$ sub-matrix of $\textbf{A}$ includes all possible $1\times r$ row vectors with the same frequency $\lambda$, then $\lambda$ is called the index of the array where the run size $n=\lambda s^r$. An $n \times m$ LHD can be considered as an OA($n,m,n,1$) with $\lambda = 1$. For every column of an OA, \cite{tang1993orthogonal} proposed to replace the $ns^{-1}$ (equal to $\lambda s^{r-1}$) positions of entry $k$ by a permutation of numbers: $(k-1)ns^{-1}+1, (k-1)ns^{-1}+2, \ldots, (k-1)ns^{-1}+ns^{-1}=kns^{-1}$, where $k=1,\ldots,s$, then the new generated design matrix would be an $n \times m$ LHD. \cite{tang1993orthogonal} showed that OALHDs can have better space-filling properties than general ones.

	\subsection{Algebraic Constructions for Orthogonal LHDs}
	\label{sec:olhd}

    Orthogonal LHDs (OLHDs) have zero pairwise correlations between any two columns and are therefore widely used in practice. There is a rich literature on the construction of OLHDs for various design sizes. In this section, we summarize several currently popular construction methods \citep{ye1998orthogonal, cioppa2007efficient, sun2010construction, tang1993orthogonal, lin2009construction, butler2001optimal}. The corresponding design sizes are summarized in Table~\ref{T1}.
	
	\begin{table}
		\caption{Summary Table of Run and Factor Sizes for Different Construction Methods.}\label{T1}
		\normalsize
		\begin{center}
			\resizebox{\textwidth}{!}{\begin{tabular}{|l|c|c|c|c|c|}
					\hline
					&Ye98&Cioppa07&Sun10&Lin09&Butler01\\
					\hline
					\multirow{2}{*}{Run ($n$)}
					& $2^m+1$ & $2^m+1$ & $r2^{c+1}$ or & $n^2$ & $n$  \\
					&         &         & $r2^{c+1}+1$  &         & \\
					\hline
					Factor ($k$) & $2m-2$ & $m+{m-1 \choose 2}$ & $2^c$ &  $2fp$ & $k \leq n-1$ \\
					\hline
					\multirow{4}{*}{Note}
					& $m$ is a & $m$ is a & $r$ and $c$ & $n^2$, $2f$ and  & $n$ is an \\
					& positive & positive & are             & $p$ are from        & odd  \\
					& integer, & integer, & positive    & OA($n^2,2f,n,2$)  & prime  \\
					& $m \geq 2$ & $m \geq 2$ & integers &  and OLHD($n$,$p$) & number \\
					\hline
			\end{tabular}}
		\end{center}
	\end{table}
	
	\cite{ye1998orthogonal} proposed a construction for OLHDs with run sizes $n=2^m+1$ and factor sizes $k=2m-2$ where $m$ is any integer greater than 2. It involves the constructions of three matrices $\textbf{M}$, $\textbf{S}$, and $\textbf{T}$. The first column in $\textbf{M}$, denoted as $\textbf{e}$, is a random permutation of $(1, 2,\ldots, 2^{m-1})$. The $2$nd to the $m$th columns in $\textbf{M}$ are calculated as $\textbf{A}_{L} \textbf{e}$, where $L=1,2,\ldots,m-1$. Each $\textbf{A}_{L}$ is defined as: $$\textbf{A}_{L}= \underbrace{\textbf{I} \otimes \ldots \otimes \textbf{I}}_\text{$m-1-L$} \otimes \underbrace{\textbf{R} \otimes \ldots \otimes \textbf{R}}_\text{$L$},$$ where $\textbf{I}$ is the $2 \times 2$ identity matrix, $\textbf{R} =  \textbf{1}\textbf{1}^T -$ $\textbf{I}$ and $\otimes$ is the Kronecker product. The last $m-2$ columns in  $\textbf{M}$ are calculated as $\textbf{A}_{i} \textbf{A}_{m-1} \textbf{e}$, where $i=1,2,\ldots,m-2$.
	Similarly, to construct the matrix $\textbf{S}$, set its first column, denoted as $\textbf{j}$, to be the $2^{m-1}\times 1$ vector of $+1$'s. The $2$nd to the $m$th columns in $\textbf{S}$, denoted as $\textbf{a}_{K}$ with $K=1,2,\ldots,m-1$ are defined as $\textbf{a}_{K}=\textbf{B}_{1} \otimes \textbf{B}_{2} \otimes \ldots \otimes \textbf{B}_{m-1}$, where all the $\textbf{B}$'s are $[1,1]^T$ except for $\textbf{B}_{m-k}=[-1,1]^T$. The last $m-2$ columns in $\textbf{S}$ are calculated as $\textbf{a}_{1} \textbf{a}_{j}$, where $j=2,\ldots,m-1$. Matrix $\textbf{T}$ is the element-wise product of $\textbf{M}$ and $\textbf{S}$, and the matrix $\textbf{X}$ is constructed by $\textbf{X}=[\textbf{T}^{T},\textbf{0},\textbf{-T}^{T}]^T$, where $\textbf{0}$ is a column vector containing $k$ zeros and $\textbf{-T} = -1 \times \textbf{T}$. For any choice of $\textbf{e}$ in $\textbf{M}$, the resulting design  $\textbf{X}$ is an OLHD.
	
	\cite{cioppa2007efficient} extended \cite{ye1998orthogonal}'s method to construct OLHDs with run size $n=2^m+1$ and factor size $k=m+ {m-1 \choose 2}$, where $m$ is any integer greater than 2. Their construction adopted the same matrix $\textbf{T}$ in \cite{ye1998orthogonal} but used different matrices $\textbf{M}$ and $\textbf{S}$. In $\textbf{M}$, for the last $m-2$ columns, instead of using $\textbf{A}_{i} \textbf{A}_{m-1} \textbf{e}$ in \cite{ye1998orthogonal}, they adopted $\textbf{A}_{i} \textbf{A}_{j} \textbf{e}$ with $i=1,\ldots,m-2$ and $j=i+1,\ldots,m-1$, which accommodates more factors. Similarly, for the matrix $\textbf{S}$, for the last $m-2$ columns, they adopted $\textbf{a}_{i} \textbf{a}_{j}$  with $i=1,\ldots,m-2$ and $j=i+1,\ldots,m-1$. Given the same number of runs, the method in \cite{cioppa2007efficient} is capable of accommodating more factors. Note that the choice of $\textbf{e}$ in constructing $\textbf{M}$ is important. If $\textbf{e}=[1, 2,\ldots, 2^{m-1}]^T$, then $\textbf{X}$ is guaranteed to be orthogonal, while a poor choice of $\textbf{e}$ may not guarantee $\textbf{X}$ to be orthogonal.
	
	\cite{sun2010construction} extended their earlier work \citep{sun2009construction} to construct OLHDs with $n=r2^{c+1}+1$ or $n=r2^{c+1}$ and $k=2^c$, where $r$ and $c$ are positive integers. Their method for constructing OLHD with $n=r2^{c+1}+1$ and $k=2^c$ consists of constructing three matrices: $S_c$, $T_c$ and $A_{r2^c\times2^c}$. The matrix $S_c$ is defined as 
	\begin{center}
		$S_c=\begin{bmatrix}
			1 & 1 \\
			1 & -1 
		\end{bmatrix}$ for $c=1$, $S_c=\begin{bmatrix}
			S_{c-1} & -S_{c-1}^{*} \\
			S_{c-1} & S_{c-1}^{*} 
		\end{bmatrix}$ for $c>1$,
	\end{center} where the operator $*$ works on any matrix with an even number of rows by multiplying the top half entries by $-1$. The matrix $T_c$ is defined as 
	\[
	T_c=\left\{
	\begin{array}{ccc}
		\begin{bmatrix}
			1 & 2 \\
			2 & -1 
		\end{bmatrix} & \mbox{ for } & c=1, \\
		\begin{bmatrix}
			T_{c-1} & -(T_{c-1}^{*}+2^{c-1}S_{c-1}^{*}) \\
			T_{c-1}+2^{c-1}S_{c-1} & T_{c-1}^{*} 
		\end{bmatrix} & \mbox{ for } & c>1.
	\end{array}
	\right.
	\]
	The matrix $A_{r2^c\times2^c}$ is defined as $A_{r2^c\times2^c}=[(T_c^{1})^{T},\ldots,(T_c^{r})^{T}]^{T}$, where $T_c^{i}=T_c+(i-1)2^cS_c$ for $i=1,\ldots,r$. The design matrix $\textbf{X}$ is given by $\textbf{X}=[A_{r2^c\times2^c}^{T},\textbf{0},-A_{r2^c\times2^c}^{T}]^T$, where $\textbf{0}$ is a column vector containing $2^c$ zeros. The construction method for OLHD with $n=r2^{c+1}$ and $k=2^c$ also relies on the $S_c$ and $T_c$. Let $H_c=T_c-S_c/2$,  $B_{r2^c\times2^c}$ is defined as $B_{r2^c\times2^c}=[(H_c^{1})^{T},\ldots,(H_c^{r})^{T}]^{T}$, where $H_c^{i}=H_c+(i-1)2^cS_c$ for $i=1,\ldots,r$. The design matrix $\textbf{X}$ is constructed as  $\textbf{X}=[B_{r2^c\times2^c}^{T},-B_{r2^c\times2^c}^{T}]^T$. The proposed method is flexible in terms of run sizes as they can be either even or odd. It accommodates more factors than that in \cite{ye1998orthogonal} but the factor sizes are restricted to be powers of two.
	
	\cite{lin2009construction} constructed OLHDs and NOLHDs with $n^2$ runs and $2fp$ factors by coupling OAs. In practice, it starts with an OLHD($n$, $p$) or NOLHD($n$, $p$), which will be coupled with an OA($n^2,2f,n,2$). Let $b_{ij}$ denote the elements of OLHD($n$, $p$), where $i=1,\ldots,n$ and $j=1,\ldots,p$, and $\textbf{A}$=OA($n^2,2f,n,2$). For $j=1,\ldots,p$, construct an $n^2 \times (2f)$ matrix $A_j$ from $\textbf{A}$ by replacing its levels $1, \ldots, n$ with $b_{ij},\ldots,b_{nj}$. Partition each $A_j$ matrix into $f$ pieces, i.e. $A_j=[A_{j1},\ldots,A_{jf}]$, where each of $A_{j1},\ldots,A_{jf}$ contains exactly two columns. Let $M_{j}=[A_{j1}V,\ldots,A_{jf}V]$, where $$V=\begin{bmatrix}
		1 & -n \\
		n & 1 
	\end{bmatrix},$$ and the design matrix  $\textbf{X}=[M_{1},\ldots,M_{p}]$. For example, an OLHD(11, 7), coupled with an OA(121,12,11,2), would yield an OLHD(121, 84). Similarly, a NOLHD(169, 168) can be generated from an NOLHD(13, 12) coupled with an OA(169,14,13,2). One advantage of their construction method is that it requires fewer runs to accommodate a large number of factors. The number of factors here must be even numbers, and the run size is restricted by the availability of the OAs.
	
	\cite{butler2001optimal} proposed a method to construct OLHDs with the run size $n$ being odd primes via the Williams transformation \citep{williams1949experimental}. When the factor size $k \leq (n-1)/2$, let $Y$ denote an $n \times k$ matrix and the elements of $Y$ are defined as 
	\[ y_{ij}=\left\{
	\begin{array}{lcll}
		ig_j+(n-1)/4 \mbox{ mod } n, & \mbox{ for } & n=1, 5, 9, \ldots,  & i.e., n \equiv 1 \mbox{ mod } 4,  \\
		ig_j+(3n-1)/4 \mbox{ mod } n, & \mbox{ for } & n=3, 7, 11, \ldots,  & i.e., n \equiv 3 \mbox{ mod } 4,  \\
	\end{array} 
	\right.\]
	where $y_{ij} \in \{0, 1, \ldots, n-1 \}$, $i=1, 2, \ldots, n$, $j=1, 2, \ldots, k$, and $g_1,g_2,\ldots,g_k$ are distinct elements in the set $\{1, 2, \ldots, (n-1)/2\}$. The design matrix would be $\textbf{X}=W(Y)$, where the Williams transformation $W$ is defined by Equation~\eqref{wt}. When $(n-1)/2+1 \leq k \leq (n-1)$, let $r=k-(n-1)/2$ and the design matrix would be partitioned as $\textbf{X}=(\textbf{X}_0,\textbf{X}_1)$, where $\textbf{X}_0$ is the $n \times (n-1)/2$ design matrix generated by the same procedure as above and $\textbf{X}_1$ is an $n \times r$ design matrix having the following elements before applying the Williams transformation:
	\begin{center}
		$x_{ij} \equiv ig_j$ mod $n$, 
	\end{center} where $x_{ij} \in \{0, 1, \ldots, n-1 \}$, $i=1, 2, \ldots, n$, $j=1, 2, \ldots, r$, and $g_1,g_2,\ldots,g_r$ are distinct elements in the set $\{1, 2, \ldots, (n-1)/2\}$. Note that $n$ can be any odd number in theory, but it is often assumed to be a prime number so that the generated designs are LHDs.

	\section{Numerical Results and Comparisons}\label{Result}

    In this section, we aim to provide guidance for practitioners on how to choose appropriate method(s) for generating optimal or near-optimal LHDs. Simulations are conducted to compare different search algorithms and algebraic constructions. To ensure fairness, all algorithms and algebraic constructions were implemented in the same software environment, namely R \citep{R}. We also make all implementations publicly available through the R package \R{LHD} (\cite{LHD}, see also \cite{wang2026rjournal}), which enables practitioners to directly apply the methods discussed in this paper.

    For each design size considered in the simulations, each search algorithm is run 20 times, and we record both the best design obtained and the average CPU time across the 20 runs. In all reported tables, a value of ``0'' indicates that the CPU time is less than one second. For all algorithms, the maximum number of iterations is set to 500,  except in the convergence analysis shown in Figure~\ref{F2}, where we extended the maximum number of iterations to 2000 to better illustrate the convergence behavior of different algorithms. In addition to the best-found results reported in the tables, Figures~\ref{F1} and~\ref{F3} present boxplots of the objective values across repeated runs. These boxplots summarize the distribution of performance, including the median, variability, and potential outliers, and thus provide insight into the stability of each algorithm.

    For the SA based algorithms (SA \citep{morris1995exploratory}, SA2008 \citep{joseph2008orthogonal}, and SLHD \citep{ba2015optimal}), the temperature decrease rate is set to $10 \%$, following \cite{morris1995exploratory} for generally good results.

    For LaPSO, let ${\it SameNumP}$ denote the number of trials used to reduce the Hamming distance with respect to the personal best for a given column within a particle, and let ${\it SameNumG}$ denote the corresponding quantity with respect to the global best. We set ${\it SameNumP} = 0$ and ${\it SameNumG} = n/4$, as recommended by \cite{chen2013optimizing}. It is worth noting that setting ${\it SameNumP} = n/2$ and ${\it SameNumG} = 0$ could yield better designs, but requires substantially more iterations (five times as many) \citep{chen2013optimizing}. The tuning parameter $p_{0}$, described in Section~\ref{LaPSo}, is set to $1/(k-1)$, also following \cite{chen2013optimizing}. We set the number of particles to $100$.

    For GA, the population size $m$ is set to $100$, matching the number of particles used in LaPSO. The mutation probability $p_{\text{mut}}$, described in Section~\ref{GA}, is set to $1/(k-1)$, consistent with the choice of $p_{0}$ in LaPSO. We aim to have equivalent settings to ensure the fairness in comparison. All parameter settings are held fixed across all design sizes. This avoids instance-specific tuning and ensures that the comparison reflects the intrinsic performance of the algorithms rather than differences in parameter alternation.
	
	\subsection{Results on Maximin Distance LHDs}\label{result1}
	
    First, we compare five search algorithms: SA \citep{morris1995exploratory}, SA2008 \citep{joseph2008orthogonal}, SLHD \citep{ba2015optimal}, LaPSO \citep{chen2013optimizing}, and GA \citep{liefvendahl2006study} together with the algebraic method FastMm \citep{wang2018optimal} for constructing maximin distance LHDs under both the $L_1$ and $L_2$ distance criteria. All of these methods and their implementation details are described in Sections~\ref{Algs} and~\ref{Constr}, respectively. We discuss the results under the $L_1$ distance in Table~\ref{T2}, and those under the $L_2$ distance in Tables~\ref{T3} and~\ref{T4}. Note that maximin $L_1$ distance designs can differ substantially from maximin $L_2$ distance designs when the design sizes are small; however, for large design sizes, the resulting designs tend to be similar.
	
	\begin{table}\caption{Minimum $\phi_{p}$ and Average CPU Time (in seconds) under the $L_1$ Distance.}\label{T2}
		\begin{center}
			\resizebox{\textwidth}{!}{
				\begin{tabular}{|c|c|c|c|c|c|c|}
					\hline
					$n \times k$&SA&SA2008&SLHD&LaPSO&GA&FastMm\\
					\hline
					$6 \times 6$&0.0874(10)&0.0874(54)&0.0883(15)&0.0856(26)&0.0856(9)&0.0856(0)\\
					\hline
					$7 \times 6$&0.0816(13)&0.0776(57)&0.0803(20)&0.0777(32)&0.0766(11)&0.0766(0)\\
					\hline
					$8 \times 8$&0.0555(26)&0.0529(127)&0.0545(37)&0.0526(62)&0.0524(19)&0.0520(0)\\
					\hline
					$9 \times 9$&0.0455(35)&0.0432(165)&0.0447(50)&0.0426(82)&0.0425(24)&0.0423(0)\\
					\hline
					$10 \times 10$&0.0380(39)&0.0358(153)&0.0380(56)&0.0354(87)&0.0353(25)&0.0353(0)\\
					\hline
					$11 \times 10$&0.0361(47)&0.0333(166)&0.0359(67)&0.0331(103)&0.0329(29)&0.0327(0)\\
					\hline
					$12 \times 12$&0.0280(70)&0.0260(290)&0.0279(99)&0.0256(157)&0.0256(43)&0.0258(0)\\
					\hline
					$13 \times 12$&0.0263(92)&0.0245(351)&0.0264(132)&0.0242(205)&0.0240(55)&0.0240(0)\\
					\hline
					$14 \times 14$&0.0212(119)&0.0197(467)&0.0211(169)&0.0194(261)&0.0194(69)&0.0193(0)\\
					\hline
				\end{tabular}
			}
		\end{center}
	\end{table}

    Table~\ref{T2} reports the minimum $\phi_{p}$ values \citep{morris1995exploratory} under the $L_1$ distance (i.e., the Manhattan distance, corresponding to $q=1$ in Equation~\eqref{E1}), along with the average CPU time for each case. Designs generated by the FastMm method are theoretically proven to be optimal when their sizes are $n \times (n-1)$ with $n$ being any odd prime, and $n \times n$ with either $2n+1$ or $n+1$ being prime \citep{wang2018optimal}. In most cases, these designs also achieve the smallest $\phi_{p}$ values in Table~\ref{T2}. Consequently, all designs constructed using the FastMm method corresponding to these theoretically covered sizes are optimal and can be obtained instantly. This includes the $12 \times 12$ case, for which $n + 1 = 13$ is prime. 
    
    We note, however, that for the $12 \times 12$ case, the $\phi_{p}$ values reported by LaPSO and GA (0.0256) are slightly smaller than the value reported for FastMm (0.0258) in Table~\ref{T2}. This apparent discrepancy arises because the theoretical optimality of FastMm is established with respect to the $L_1$ distance criterion, $d_1(\textbf{X})$, defined at the beginning of Section~\ref{OC}, whereas Table~\ref{T2} reports the $\phi_{p}$ criterion. Although $\phi_{p}$ is commonly used as a surrogate measure for constructing maximin designs, it depends on the entire set of pairwise distances rather than only the minimum distance. Consequently, for a finite value of $p$, minimizing $\phi_{p}$ does not necessarily induce the same ranking as maximizing $d_1(\textbf{X})$. For example, a design may have a slightly smaller minimum distance but a more favorable overall distance distribution, resulting in a smaller $\phi_{p}$ value. We verified that, for the $12 \times 12$ case, the design generated by FastMm has $d_1(\textbf{X}) = 50$, whereas the best designs found by GA and LaPSO have $d_1(\textbf{X}) = 49$. Therefore, the FastMm design remains optimal with respect to the maximin $L_1$ distance criterion, consistent with the theoretical result, because it achieves a larger $L_1$ distance.
    
    Among the five search algorithms, GA achieves the smallest $\phi_{p}$ values while also requiring the lowest average CPU time, and in some cases identifies the same optimal designs as FastMm. LaPSO performs comparably to GA in terms of $\phi_{p}$ values but requires substantially longer CPU time. Overall, for the special design sizes covered by \cite{wang2018optimal}, the algebraic construction FastMm is recommended. When a search algorithm is required, GA is the preferred choice.
	
	\begin{table}\caption{Minimum $\phi_{p}$ and Average CPU Time (in seconds) with Run Size Between 4 and 14 under the $L_2$ Distances.}\label{T3}
		\begin{center}
			\resizebox{\textwidth}{!}{
				\begin{tabular}{|l|c|c|c|c|c|c|c|c|c|c|}
					\cline{1-5}\cline{7-11}
					&$k$&{2}&{3}&{4}& &&$k$&{2}&{3}&{4}\\
					
					\cline{1-5}\cline{7-11} 				
					$n$& &Min(CPU)&Min(CPU)&Min(CPU)& &$n$& &Min(CPU)&Min(CPU)&Min(CPU)\\
					
					\cline{1-5}\cline{7-11}
					\multirow{6}{*}{4}&SA&0.4906(5)&0.4113(6)&0.3137(6)&&
					\multirow{6}{*}{8}&SA&0.3961(12)&0.2657(13)&0.2064(14)\\
					&SA2008&0.4906(13)&0.4113(21)&0.3137(30)&&                  &SA2008&0.3961(24)&0.2653(29)&0.2046(37)\\
					&SLHD&0.4906(9)&0.4113(9)&0.3137(9)&&                  &SLHD&0.3961(18)&0.2647(19)&0.2043(21)\\
					&LaPSO&0.4906(13)&0.4113(14)&0.3137(15)&&                  &LaPSO&0.3961(26)&0.2556(29)&0.1907(32)\\
					&GA&0.4906(6)&0.4113(6)&0.3137(7)&&
					&GA&0.3961(8)&0.2556(9)&0.1907(10)\\
					&FastMm&0.4906(0)&0.4290(0)&0.3469(0)&&                  &FastMm&0.4907(0) &0.3501(0) &0.2049(0)\\
					\cline{1-5}\cline{7-11}
					
					\multirow{6}{*}{5}&SA&0.4907(5)&0.3351(6)&0.2715(7)&&
					\multirow{6}{*}{10}&SA&0.3753(21)&0.2430(23)&0.1882(24)\\
					&SA2008&0.4907(13)&0.3451(17)&0.2715(26)&&                  &SA2008&0.3631(41)&0.2367(48)&0.1820(57)\\
					&SLHD&0.4907(9)&0.3351(9)&0.2715(11)&&                  &SLHD&0.3696(31)&0.2437(32)&0.1849(34)\\
					&LaPSO&0.4907(13)&0.3351(14)&0.2715(17)&&                  &LaPSO&0.3631(46)&0.2257(49)&0.1736(53)\\
					&GA&0.4907(5)&0.3351(6)&0.2715(7)&&
					&GA&0.3631(14)&0.2270(14)&0.1732(15)\\
					&FastMm&0.4907(0)&0.4292(0)&0.2844(0)&&                  &FastMm&0.4816(0) &0.2758(0) &0.1844(0)\\
					\cline{1-5}\cline{7-11}
					
					\multirow{6}{*}{6}&SA&0.4821(8)&0.2974(9)&0.2421(10)&&
					\multirow{6}{*}{12}&SA&0.3602(24)&0.2279(27)&0.1721(30)\\
					&SA2008&0.4821(18)&0.2974(25)&0.2414(33)&&                  &SA2008&0.3338(44)&0.2115(52)&0.1651(61)\\
					&SLHD&0.4821(12)&0.2974(13)&0.2414(14)&&                  &SLHD&0.3567(35)&0.2275(39)&0.1739(43)\\
					&LaPSO&0.4821(18)&0.2974(21)&0.2414(23)&&                  &LaPSO&0.3338(51)&0.2044(57)&0.1524(64)\\
					&GA&0.4821(7)&0.2974(8)&0.2414(8)&&
					&GA&0.3338(14)&0.2024(16)&0.1534(18)\\
					&FastMm&0.4907(0)&0.3021(0)&0.2577(0)&&                  &FastMm&0.4696(0) &0.2475(0) &0.1608(0)\\
					\cline{1-5}\cline{7-11}
					
					\multirow{6}{*}{7}&SA&0.3961(11)&0.2758(12)&0.2237(13)&&
					\multirow{6}{*}{14}&SA&0.3496(37)&0.2187(44)&0.1622(48)\\
					&SA2008&0.3961(23)&0.2811(30)&0.2181(38)&&                  &SA2008&0.3313(66)&0.2030(82)&0.1544(92)\\
					&SLHD&0.3961(16)&0.2758(17)&0.2197(18)&&                  &SLHD&0.3478(54)&0.2145(63)&0.1600(68)\\
					&LaPSO&0.3961(24)&0.2758(26)&0.2162(29)&&                  &LaPSO&0.3256(77)&0.1865(92)&0.1407(99)\\
					&GA&0.3961(8)&0.2758(9)&0.2162(10)&&
					&GA&0.3189(21)&0.1872(25)&0.1407(27)\\
					&FastMm&0.4907(0)&0.3014(0)&0.2329(0)&&                  &FastMm&0.4908(0) &0.2259(0) &0.1677(0)\\
					\cline{1-5}\cline{7-11}
			\end{tabular}}
		\end{center}
	\end{table}

    Since the $L_2$ distance (i.e., the Euclidean distance, corresponding to $q=2$ in Equation~\eqref{E1}) is the most popular choice in practice, we focus on maximin $L_2$ distance designs in the following discussion. Table~\ref{T3} presents the results for designs with run sizes ranging from 4 to 14. When $k = 2$ with $n = 4$ or $n = 5$, all methods achieve the same minimum $\phi_{p}$ value, while FastMm requires the least CPU time. For $k = 2$ with $n = 6, 7,$ or $8$, all search algorithms yield the same minimum $\phi_{p}$, and GA has the lowest CPU time. When $k = 2$ with $n = 10$ or $n = 12$, both LaPSO and GA attain the same minimum $\phi_{p}$. For the $14 \times 2$ design, GA achieves the smallest minimum $\phi_{p}$. When $k = 3$ with $n \leq 7$ and $k = 4$ with $n \leq 6$, nearly all search algorithms produce the same minimum $\phi_{p}$, with GA again requiring the least CPU time. In all remaining cases, GA and LaPSO achieve similar minimum $\phi_{p}$ values and outperform the other search algorithms; however, GA consistently requires the least computational time.

    In Table~\ref{T4}, we consider designs following the rule of thumb run size ($n = 10k$) commonly used in computer experiments \citep{chapman1994arctic, jones1998efficient, loeppky2009choosing, harari2018computer}. Table~\ref{T4} shows that as both $n$ and $k$ increase, the GA tends to require the least CPU time among the five search algorithms. In most cases, GA and LaPSO outperform the other methods in terms of the $\phi_{p}$ values, with GA often yielding better results than LaPSO. LaPSO generally incurs the highest CPU time. Therefore, we recommend GA when computational resources are limited.
	
	\begin{table}\caption{Minimum $\phi_{p}$ and Average CPU Time (in minutes) with the Rule of Thumb Sizes under the $L_2$ Distances.}\label{T4}
		\begin{center}
			\resizebox{\textwidth}{!}{\begin{tabular}{|c|c|c|c|c|c|c|c|}
					\hline
					$n \times k$&{20 $\times$ 2}&{30 $\times$ 3}&{40 $\times$ 4}&{50 $\times$ 5}&{60 $\times$ 6}&{70 $\times$ 7}&{80 $\times$ 8}\\
					\hline
					&Min(CPU)&Min(CPU)&Min(CPU)&Min(CPU)&Min(CPU)&Min(CPU)&Min(CPU)\\
					\hline
					
					SA     &0.3344(1.4)&0.1707(3.3)&0.0999(7.9)&0.0678(13.7)&0.0499(22.0)&0.0380(37.6)&0.0292(55.4)\\
					SA2008 &0.3021(2.4)&0.1471(5.3)&0.0881(12.3)&0.0589(21.2)&0.0428(33.1)&0.0332(36.9)&0.0266(79.6)\\
					SLHD   &0.3070(2.1)&0.1643(4.8)&0.1002(11.2)&0.0685(19.6)&0.0490(31.0)&0.0373(43.2)&0.0297(76.3)\\
					LaPSO  &0.2849(3.0)&0.1283(6.8)&0.0753(16.0)&0.0512(27.8)&0.0378(44.9)&0.0292(75.9)&0.0236(99.4)\\
					GA     &0.2830(0.8)&0.1252(1.7)&0.0740(4.0) &0.0499(7.0)&0.0371(11.1)&0.0287(18.9)&0.0232(28.0)\\
					FastMm &0.4905(0)  &0.1565(0)  &0.0895(0)   &0.0634(0)   &0.0450(0)   &0.0342(0)&0.0280(0)\\

					\hline
			\end{tabular}}
		\end{center}
	\end{table}

	\begin{figure}\caption{Boxplots of $\phi_{p}$ Values from Different Algorithms with the Rule of Thumb Sizes.}\label{F1}
		\begin{center}
			\resizebox{\textwidth}{!}{\begin{tabular}{cc}
					\includegraphics[scale=0.45]{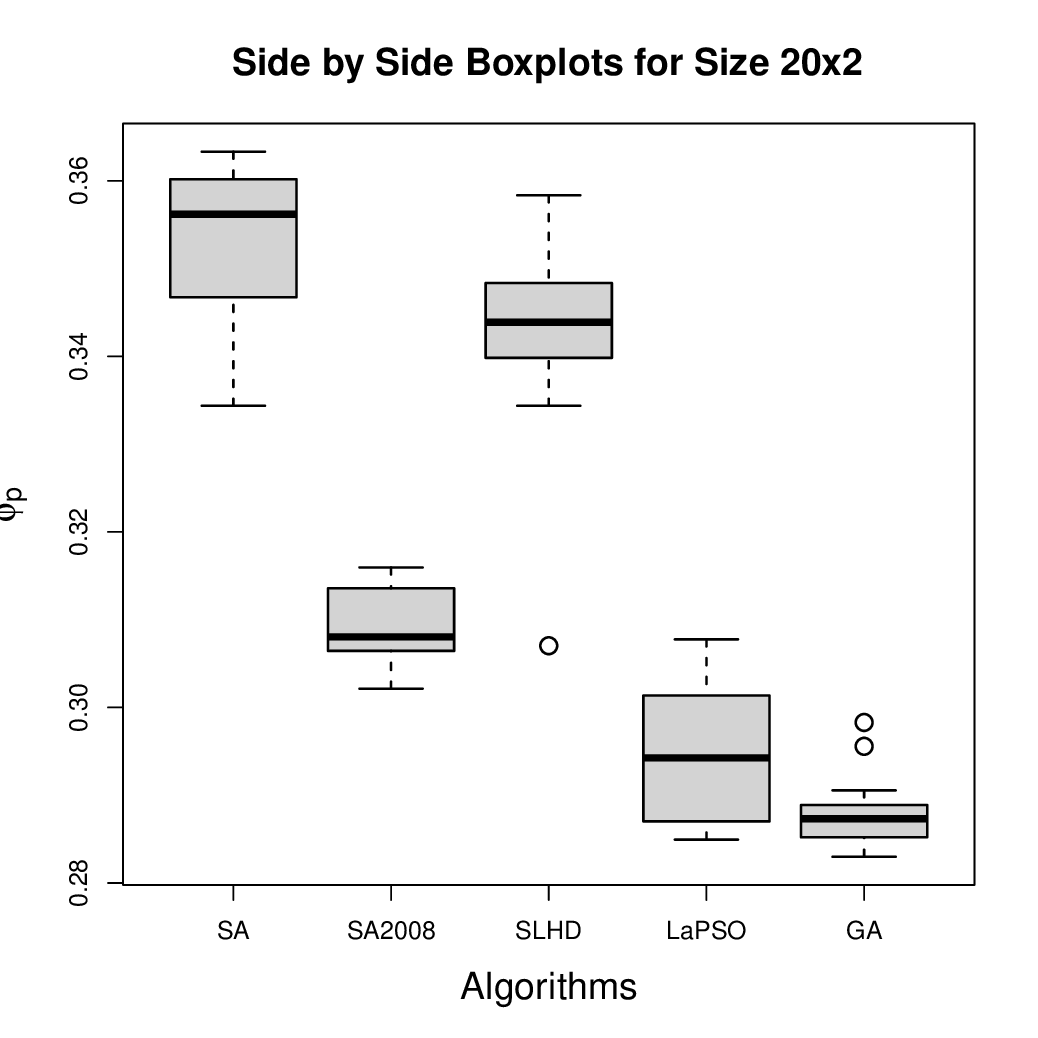}&\includegraphics[scale=0.45]{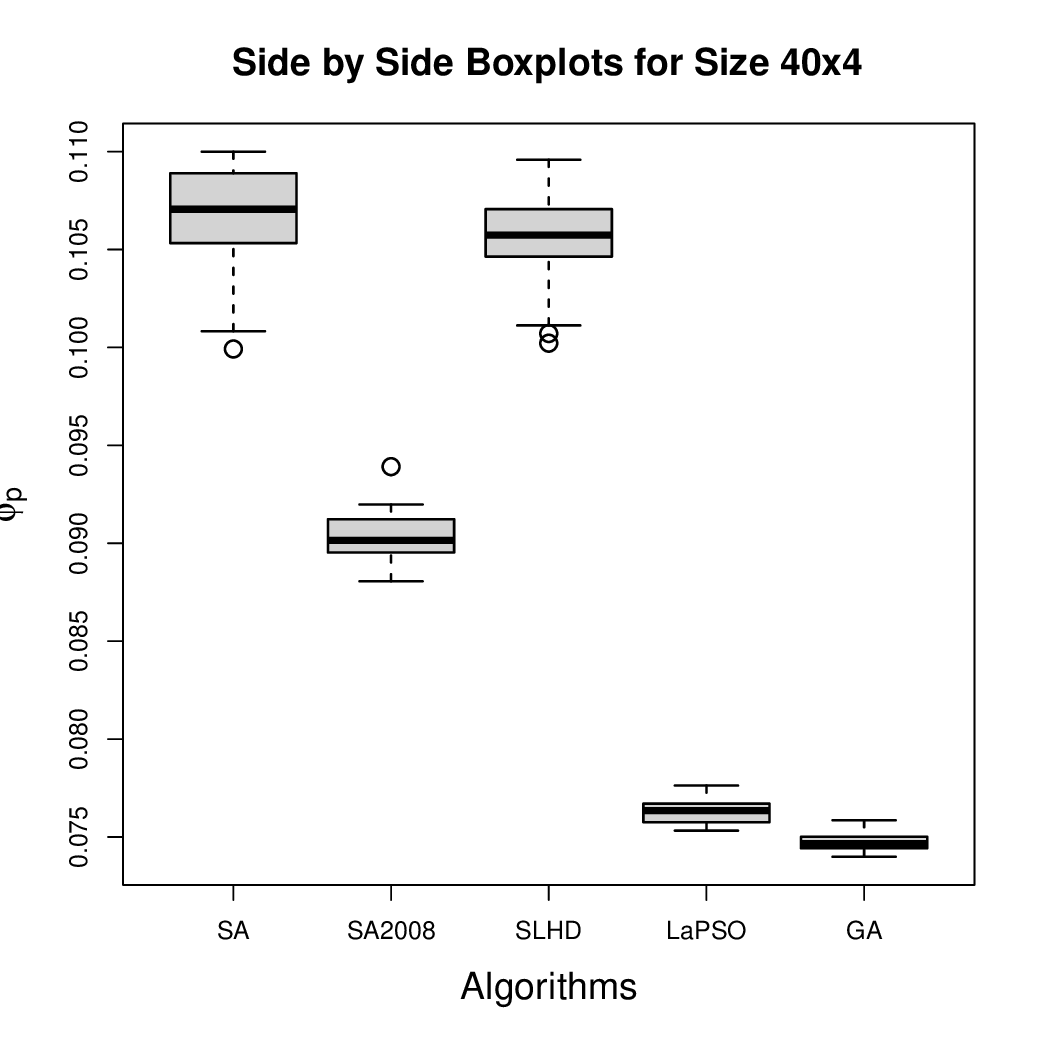} \\
					\includegraphics[scale=0.45]{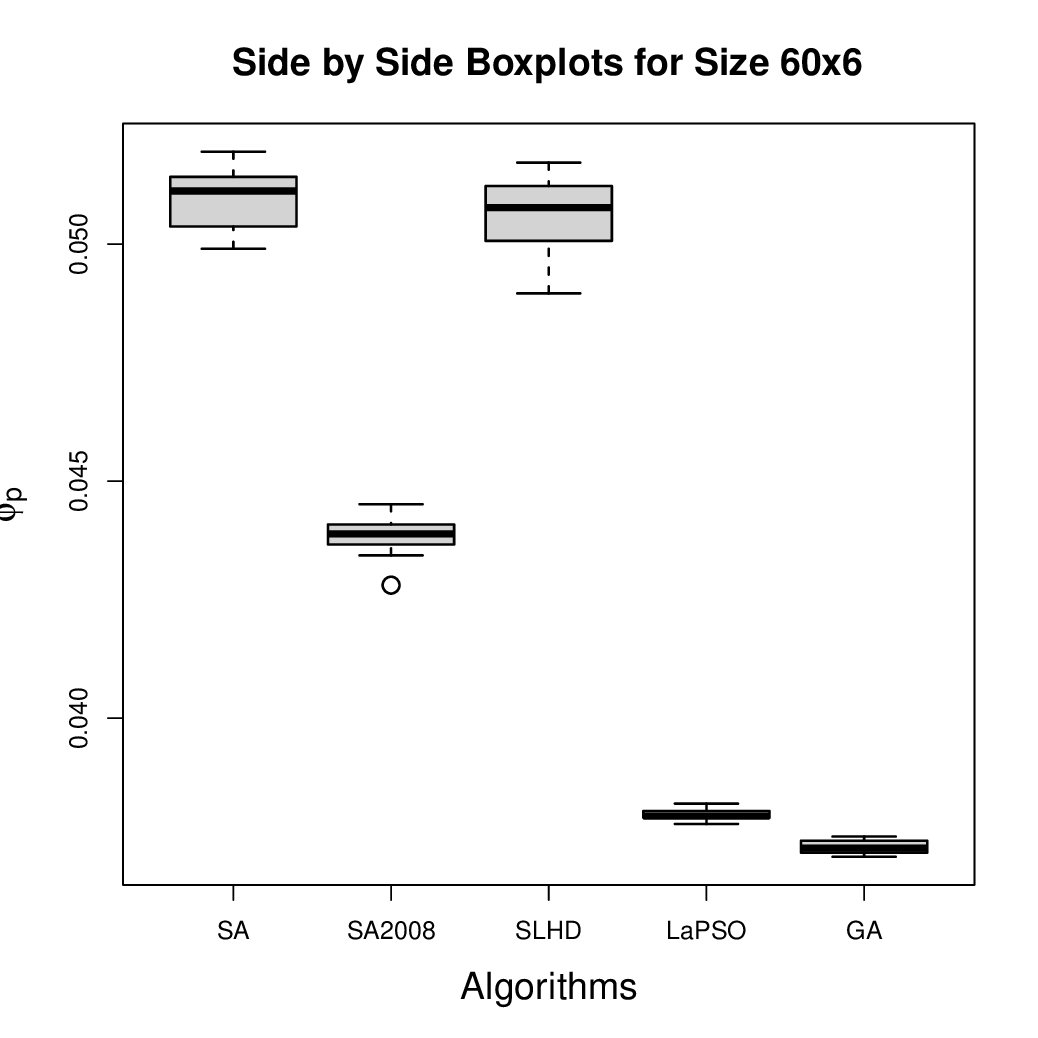}&\includegraphics[scale=0.45]{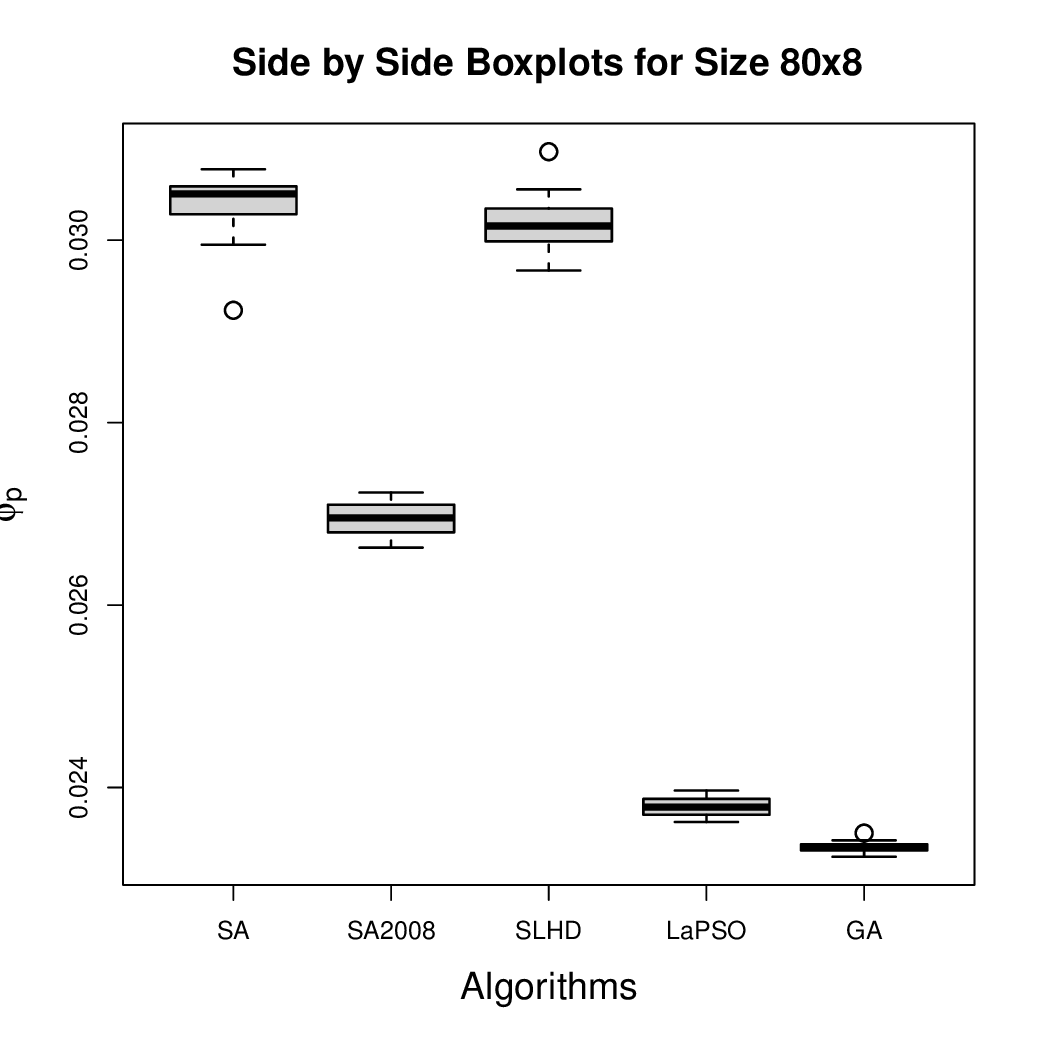} \\
			\end{tabular}}
		\end{center}
	\end{figure}

    We further present boxplots of the $\phi_{p}$ values for selected designs with rule of thumb run sizes in Figure~\ref{F1}. For the case of $n = 20$ and $k = 2$ (the top-left panel of Figure~\ref{F1}), GA yields the smallest median $\phi_{p}$ value, while LaPSO appears less stable than the other algorithms. In the remaining three panels, GA and LaPSO outperform the other methods. In most cases, the relative performance patterns observed in terms of the best-found values are consistent with those of the median values, suggesting that the conclusions drawn from the tables are robust. Moreover, algorithms with tighter interquartile ranges, such as GA, tend to exhibit more stable performance across runs. As the design sizes increase, the advantage of using GA becomes increasingly obvious.

	\begin{figure}\caption{Convergence Curves for Different Algorithms with the Rule of Thumb Sizes.}\label{F2}
		\begin{center}
			\resizebox{\textwidth}{!}{\begin{tabular}{cc}
					\includegraphics[scale=0.45]{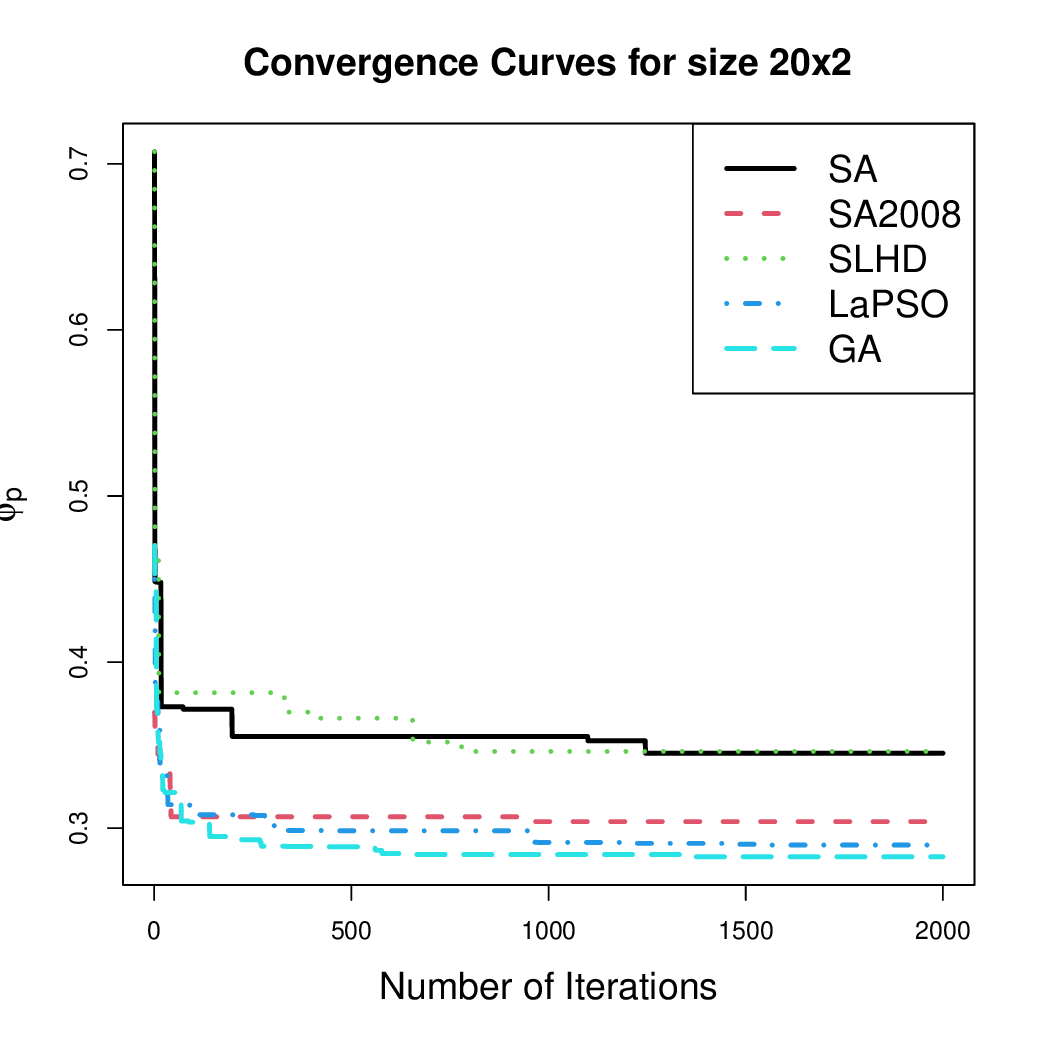}&\includegraphics[scale=0.45]{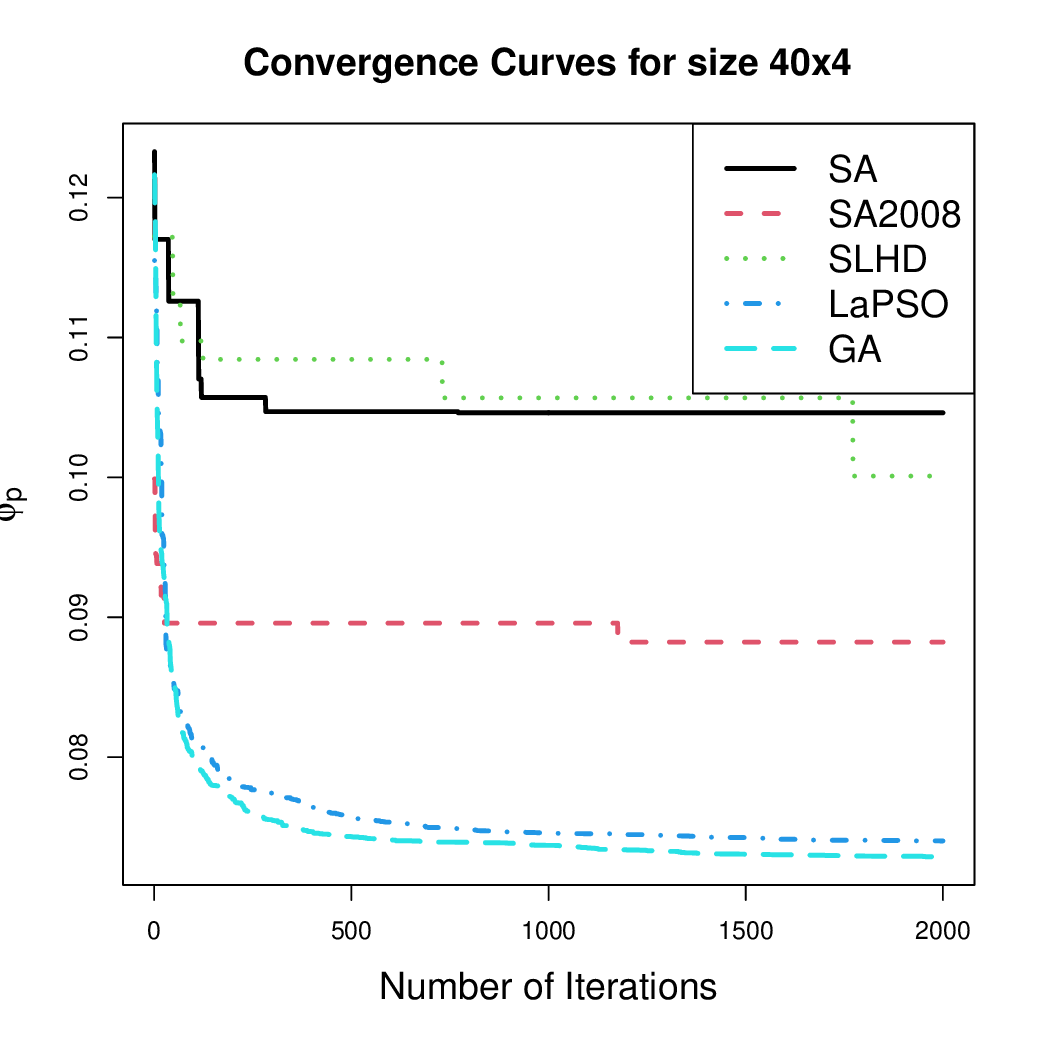} \\
					\includegraphics[scale=0.45]{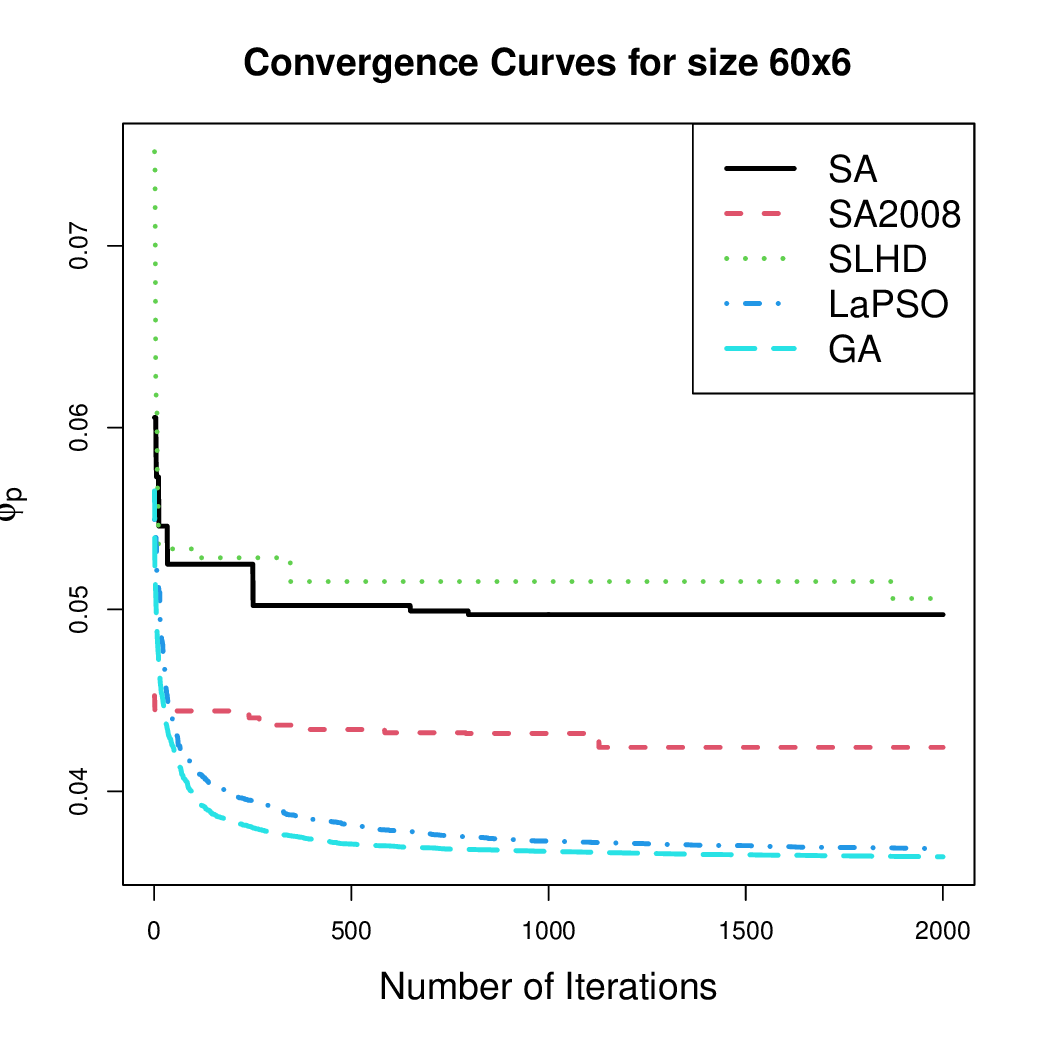}&\includegraphics[scale=0.45]{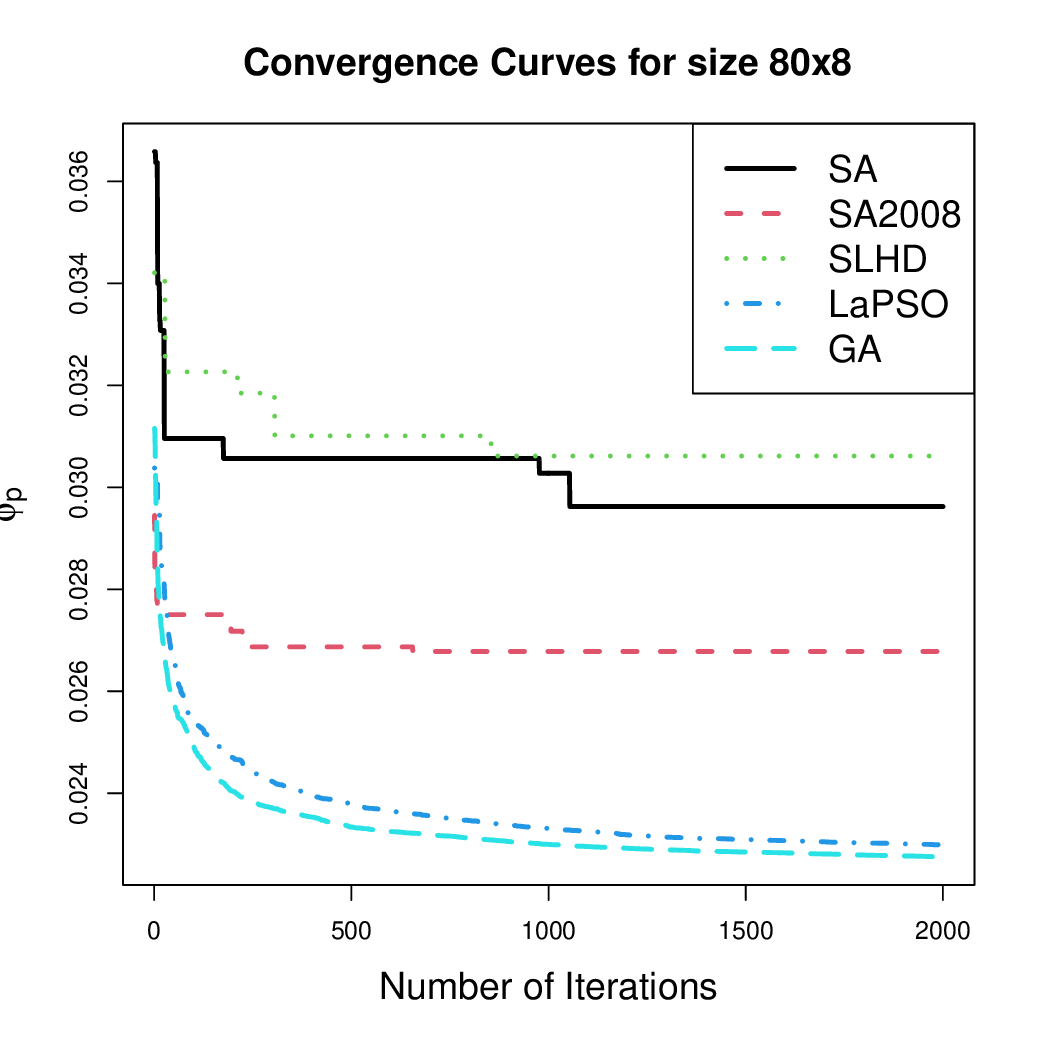} \\
			\end{tabular}}
		\end{center}
	\end{figure}

    Figure~\ref{F2} illustrates the convergence behavior of each algorithm for different design sizes. For the case of $n = 20$ and $k = 2$ (the top-left panel of Figure~\ref{F2}), SA converges around the 1300th iteration, whereas the remaining algorithms all converge before the 500th iteration. For the $40 \times 4$ and $60 \times 6$ cases (the top-right and bottom-left panels of Figure~\ref{F2}), SLHD continues to show improvements until approximately the 2000th iteration, while the other algorithms stop improving around the 1200th iteration. Overall, GA converges faster than all other methods while achieving strong performance, with LaPSO ranking second. Note that GA generates new candidate designs by exchanging entire columns, whereas the other methods rely on element-wise exchanges. Element-wise operations are effective when $n$ and $k$ are relatively small; however, as the size of designs increases, the number of elements grows exponentially, and element-wise exchange strategies may become less efficient.

    We note that the computational meaning of an ``iteration’’ differs across algorithms. For population-based algorithms (i.e., GA and LaPSO), one iteration involves updating multiple candidate designs, whereas for SA-based algorithms, each iteration updates a single design. Accordingly, the convergence curves should be interpreted qualitatively as indicators of stabilization behavior rather than as measures of computational efficiency.

	\subsection{Results on Maximum Projection LHDs}\label{result2}

    \cite{joseph2015maximum} adopted a simulated annealing algorithm to identify MaxPro LHDs, which is implemented in the R package \R{MaxPro} \citep{MaxPro} (denoted as “MaxPro” in the tables). Here, we compare this approach with the LaPSO and GA algorithms implemented in the R package \R{LHD} \citep{LHD}. The objective is to identify MaxPro LHDs by minimizing the objective function $\psi$ defined in Equation~\eqref{E2}. For each design size, we run each algorithm 20 times and report the best results (i.e., the minimum $\psi$ values) along with the average CPU time in Tables~\ref{T6} and \ref{T7}. For the algorithm from the \R{MaxPro} package, we set the number of iterations to 500 and keep all other input arguments at their default settings. For both LaPSO and GA, all parameter settings follow those specified at the beginning of Section~\ref{Result}.

    Note that the \R{MaxPro} package was developed in C++ using Rcpp framework, while the \R{LHD} package was developed entirely in the standard R environment. Consequently, \R{MaxPro} achieves substantially lower CPU times in all cases. Therefore, the reported CPU times are provided for information and reference purposes only.

	\begin{table}\caption{Minimum $\psi$ and Average CPU Time (in seconds) with Run Size Between 4 and 14.}\label{T6}
		\begin{center}
			\resizebox{\textwidth}{!}{
				\begin{tabular}{|l|c|c|c|c|c|l|c|c|c|c|}
					\cline{1-5}\cline{7-11}
					&$k$&{2}&{3}&{4}& &&$k$&{2}&{3}&{4}\\
					
					\cline{1-5}\cline{7-11} 				
					$n$& &Min(CPU)&Min(CPU)&Min(CPU)& &$n$& &Min(CPU)&Min(CPU)&Min(CPU)\\
					
					\cline{1-5}\cline{7-11}
					\multirow{3}{*}{4}&MaxPro&0.4513(0)&0.4705(0)&0.4454(0)&&
					\multirow{3}{*}{8}&MaxPro&0.2240(0)&0.2072(0)&0.2021(0)\\
					&LaPSO&0.4513(5)&0.4705(5)&0.4454(6)&&                  &LaPSO&0.2240(8)&0.1737(9)&0.1763(11)\\
					&GA&0.4513(3)&0.4705(3)&0.4454(4)&&
					&GA&0.2240(4)&0.1737(4)&0.1763(5)\\
					\cline{1-5}\cline{7-11}
					
					\multirow{3}{*}{5}&MaxPro&0.3771(0)&0.3561(0)&0.3382(0)&&
					\multirow{3}{*}{10}&MaxPro&0.1800(0)&0.1682(0)&0.1625(0)\\
					&LaPSO&0.3771(5)&0.3561(6)&0.3382(7)&&                  &LaPSO&0.1685(9)&0.1404(11)&0.1307(13)\\
					&GA&0.3771(3)&0.3561(4)&0.3382(4)&&
					&GA&0.1685(4)&0.1386(5)&0.1304(5)\\
					\cline{1-5}\cline{7-11}
					
					\multirow{3}{*}{6}&MaxPro&0.3154(0)&0.2633(0)&0.2724(0)&&
					\multirow{3}{*}{12}&MaxPro&0.1461(0)&0.1371(0)&0.1322(0)\\
					&LaPSO&0.3154(6)&0.2633(6)&0.2551(7)&&                  &LaPSO&0.1330(9)&0.1115(11)&0.1010(13)\\
					&GA&0.3154(3)&0.2633(4)&0.2551(4)&&
					&GA&0.1330(4)&0.1120(4)&0.0990(5)\\
					\cline{1-5}\cline{7-11}
					
					\multirow{3}{*}{7}&MaxPro&0.2511(0)&0.2300(0)&0.2301(0)&&
					\multirow{3}{*}{14}&MaxPro&0.1299(0)&0.1174(0)&0.1077(0)\\
					&LaPSO&0.2511(6)&0.2184(7)&0.2113(8)&&                  &LaPSO&0.1149(15)&0.0926(18)&0.0826(20)\\
					&GA&0.2511(3)&0.2184(4)&0.2113(5)&&
					&GA&0.1145(6)&0.0914(6)&0.0819(7)\\
					\cline{1-5}\cline{7-11}
			\end{tabular}}
		\end{center}
	\end{table}

    In Table~\ref{T6}, when the design sizes are small (i.e., $k = 2$ with $n \leq 8$, $k = 3$ with $n \leq 6$, and $k = 4$ with $n \leq 5$), all three algorithms yield the same minimum $\psi$ values, while the MaxPro method has the lowest CPU time. For the remaining design sizes (except for the case of $12 \times 3$), the GA produces the best results. The LaPSO also outperforms the MaxPro method, with $\psi$ values close to those obtained by the GA. In Table~\ref{T7}, we present additional results for designs with rule of thumb run sizes. For relatively large design sizes, the GA consistently identifies designs with the smallest $\psi$ values. The MaxPro method requires the least CPU time, whereas the LaPSO generally outperforms MaxPro in terms of design quality but incurs the highest CPU time among the three methods.
	
	\begin{table}\caption{Minimum $\psi$ and Average CPU Time (in minutes) with the Rule of Thumb Sizes.}\label{T7}
		\begin{center}
			\resizebox{\textwidth}{!}{\begin{tabular}{|c|c|c|c|c|c|c|c|}
					\hline
					$n \times k$&{20 $\times$ 2}&{30 $\times$ 3}&{40 $\times$ 4}&{50 $\times$ 5}&{60 $\times$ 6}&{70 $\times$ 7}&{80 $\times$ 8}\\
					\hline
					&Min(CPU)&Min(CPU)&Min(CPU)&Min(CPU)&Min(CPU)&Min(CPU)&Min(CPU)\\
					\hline
					
					MaxPro&0.0870(0)  &0.0510(0)  &0.0335(0)  &0.0227(0)  &0.0163(0)  &0.0132(0)   &0.0101(0)\\
					
					LaPSO  &0.0751(0.4)&0.0343(0.8)&0.0204(2.2)&0.0138(3.7)&0.0098(4.4)&0.0075(6.7)&0.0059(9.4)\\
					
					GA     &0.0749(0.1)&0.0335(0.2)&0.0193(0.5)&0.0128(0.9)&0.0093(1.1)&0.0071(1.6) &0.0056(2.3)\\

					\hline
			\end{tabular}}
		\end{center}
	\end{table}

	\begin{figure}\caption{Boxplots of $\psi$ values from Different Algorithms with the Rule of Thumb Design Sizes.}\label{F3}
		\begin{center}
			\resizebox{\textwidth}{!}{\begin{tabular}{cc}
					\includegraphics[scale=0.45]{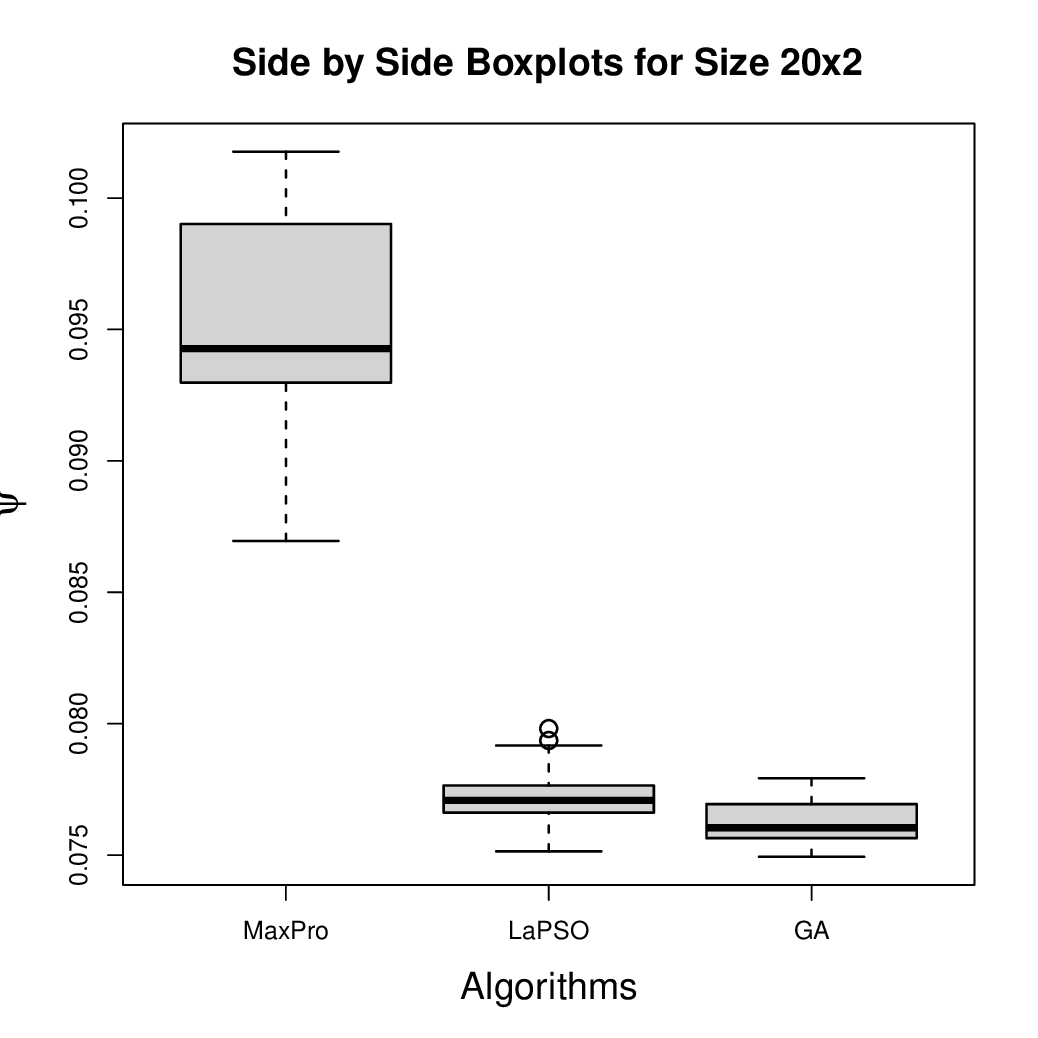}&\includegraphics[scale=0.45]{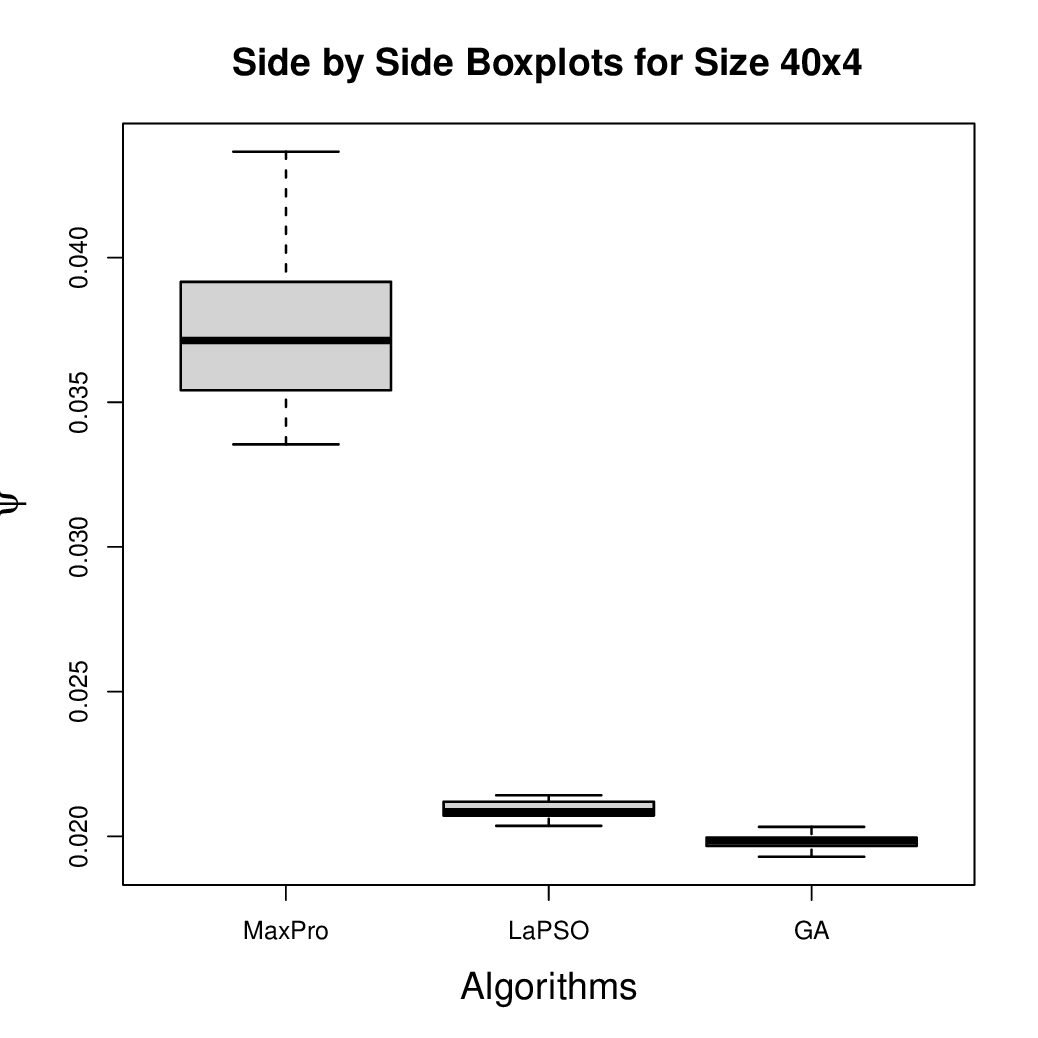} \\
					\includegraphics[scale=0.45]{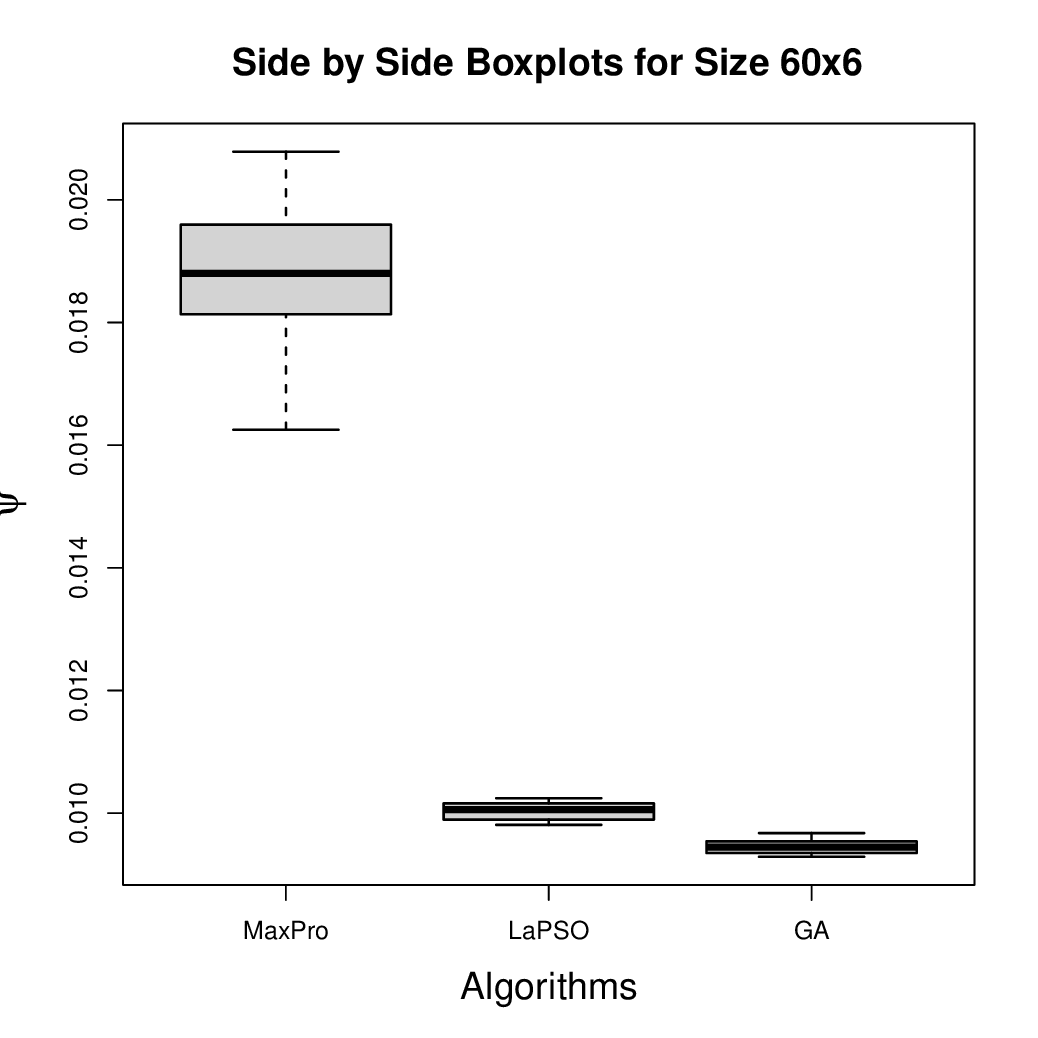}&\includegraphics[scale=0.45]{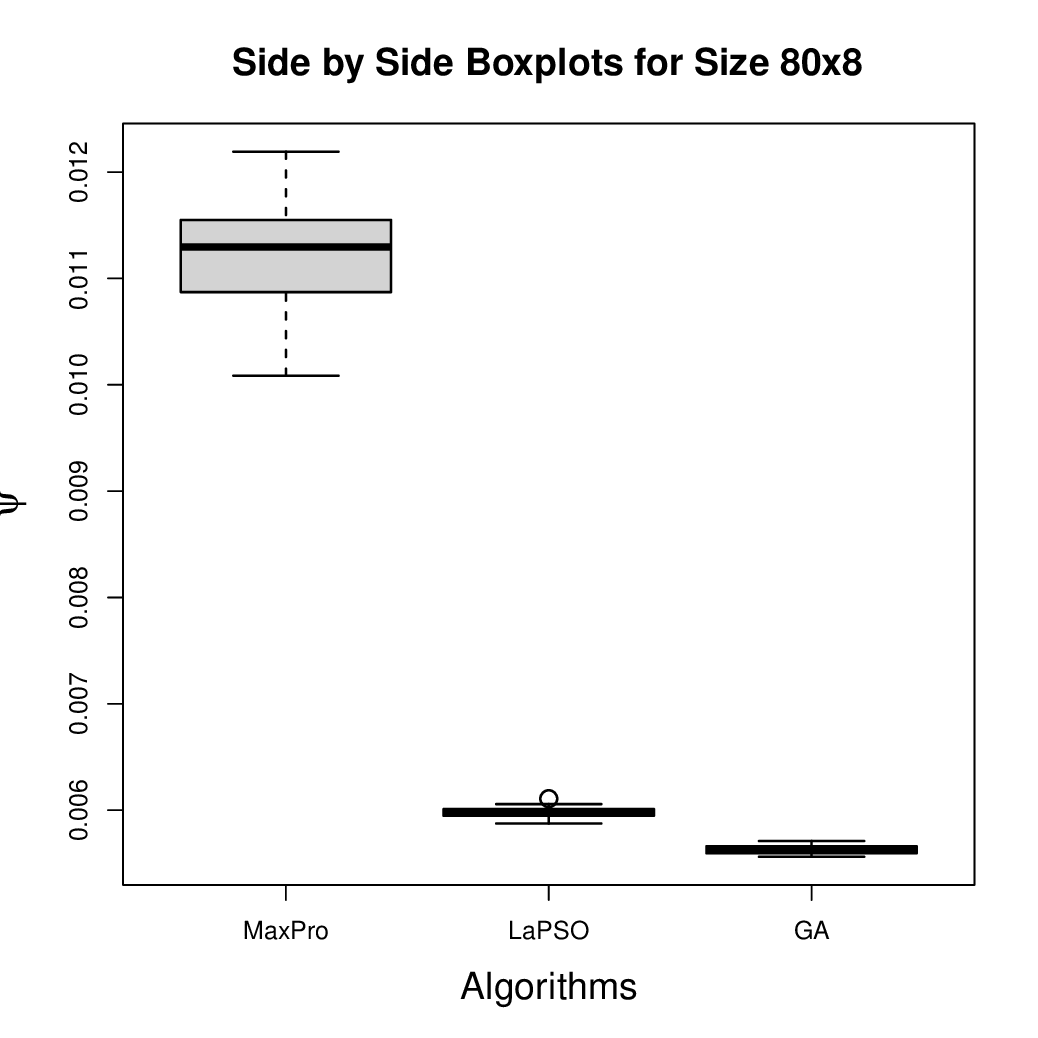} \\
			\end{tabular}}
		\end{center}
	\end{figure}

    Figure~\ref{F3} presents boxplots of the $\psi$ values for several cases with rule of thumb design sizes. In the upper left panel of Figure~\ref{F3} ($k = 2$ and $n = 20$), the MaxPro method appears less stable than the others, as indicated by its more dispersed boxplot. The GA identifies the best-found design and exhibits the greatest stability. In the remaining three panels of Figure~\ref{F3}, the GA again outperforms the other methods. As the design size increases, the advantage of the GA becomes more obvious. Generally speaking, GA is recommended for constructing MaxPro LHDs of moderate to large sizes, offering reasonable CPU times. When computational time is the primary concern, the MaxPro method is preferred.
	
	\subsection{Results on Orthogonal and Nearly Orthogonal LHDs}\label{result3}

    In this subsection, we aim to identify orthogonal LHDs (OLHDs) and nearly orthogonal LHDs (NOLHDs) by minimizing the ave$(|q|)$ or max$|q|$ criteria defined in Equation~\eqref{E3}. Tables~\ref{T8} to \ref{T10} present results for selected designs with run sizes up to $2^8$. In these tables, “CM” indicates whether an algebraic construction method is available for the given design size. If “CM” is “Yes,” the corresponding OLHD can be obtained using the construction method; otherwise, search algorithms are required. For a summary of available algebraic constructions across different design sizes, please refer to Table~\ref{T1} in Section~\ref{sec:olhd}. For all the algorithms considered in this subsection, the parameter settings follow those specified at the beginning of Section~\ref{Result}.
	
	\begin{table}\caption{Minimum ave$(|q|)$, max$|q|$, and Average CPU Time (in seconds) with Run Size Below $2^4$.}\label{T8}
		\normalsize
		\begin{center}
			\begin{tabular}{|l|c|c|c|c|c|c|c|}
				\hline
				$n\times k$&Criteria&{SA}&{SA2008}&{SLHD}&{LaPSO}&{GA}&CM\\
				\hline
				\multirow{3}{*}{$7\times 4$}
				& ave$(|q|)$ &0.0298&0.0060&0.0238&0.0060&0.0179&\\
				& max$|q|$   &0.0714&0.0357&0.0714&0.0357&0.0357&No\\
				& CPU        &22&47&30&47&14&\\
				\hline
				\multirow{3}{*}{$8\times 4$}
				& ave$(|q|)$ &0.0198&0.0000&0.0317&0.0079&0.0079&\\
				& max$|q|$   &0.0476&0.0000&0.0476&0.0238&0.0238&Yes\\
				& CPU        &14&34&20&32&10&\\
				\hline
				\multirow{3}{*}{$9\times 4$}
				& ave$(|q|)$ &0.0278&0.0000&0.0194&0.0083&0.0000&\\
				& max$|q|$   &0.0500&0.0000&0.0500&0.0167&0.0000&Yes\\
				& CPU        &14&36&21&32&10&\\
				\hline
				\multirow{3}{*}{$10\times 4$}
				& ave$(|q|)$ &0.0283&0.0061&0.0283&0.0061&0.0061&\\
				& max$|q|$   &0.0545&0.0061&0.0545&0.0061&0.0061&No\\
				& CPU        &22&56&32&48&14&\\
				\hline
				\multirow{3}{*}{$10\times 6$}
				& ave$(|q|)$ &0.0877&0.0174&0.1030&0.0246&0.0166&\\
				& max$|q|$   &0.2121&0.0424&0.2121&0.0545&0.0303&No\\
				& CPU        &64&135&88&136&37&\\
				\hline
				\multirow{3}{*}{$12\times 6$}
				& ave$(|q|)$ &0.0932&0.0121&0.0769&0.0154&0.0145&\\
				& max$|q|$   &0.1748&0.0350&0.1818&0.0280&0.0280&No\\
				& CPU        &63&144&89&137&37&\\
				\hline
				\multirow{3}{*}{$14\times 6$}
				& ave$(|q|)$ &0.0769&0.0086&0.0769&0.0125&0.0092&\\
				& max$|q|$   &0.1473&0.0198&0.1516&0.0242&0.0198&No\\
				& CPU        &44&116&65&96&26&\\
				\hline
			\end{tabular}
		\end{center}
	\end{table}

    In Table~\ref{T8}, algebraic constructions are available for two cases ($8 \times 4$ and $9 \times 4$). SA2008 can also identify these OLHDs, with both ave$(|q|)$ and max$|q|$ equal to 0. For the remaining cases, SA2008 is capable of identifying NOLHDs with very small ave$(|q|)$ and max$|q|$ values. In general, SA2008 outperforms all other search algorithms, except for the $10 \times 6$  and $12 \times 6$ cases, in which GA performs better. In Table~\ref{T9}, we further compare SA2008 and GA for larger design sizes. Note that the SA, the SLHD and the LaPSO  perform clearly worse than the SA2008 and the GA in our study, and we do not include them in Tables~\ref{T9} to \ref{T10} for conciseness. When $n \geq 16$ and $k \geq 6$, GA provides the best results, achieving very small ave$(|q|)$ and max$|q|$ values. Moreover, its CPU time is substantially lower than that of SA2008.

    In Table~\ref{T10}, we present additional cases with even larger run sizes. Only a few of these cases have available algebraic constructions. The designs generated by GA achieve max$|q|$ values below 0.025 in all cases while requiring reasonable CPU time. Moreover, in Table~\ref{T11}, we consider cases with rule of thumb design sizes, comparing all five algorithms implemented in the R package \R{LHD}. For the $20 \times 2$ design, all five algorithms successfully identify OLHDs, with GA and SA achieving the lowest CPU times. For cases with $k \geq 3$, GA achieves the smallest max$|q|$ values (below 0.0024) and requires the least CPU time in all instances. In general, SA2008 performs best for small designs (i.e., $n < 16$), while GA is recommended for moderate and large designs. This conclusion aligns with the discussion in Section~\ref{result1}: element-wise operations are effective when $n$ and $k$ are relatively small, while GA generates new candidate designs by exchanging entire columns, which is less affected as design sizes increase.
	
	\begin{table}\caption{Minimum ave$(|q|)$, max$|q|$, and Average CPU Time (in seconds) with Run Size Between $2^4$ and $2^5-1$.}\label{T9}
		\begin{center}
			\resizebox{\textwidth}{!}{\begin{tabular}{|l|c|c|c|c|c|l|c|c|c|c|}
					\cline{1-5}\cline{7-11}
					$n\times k$&Criteria&{SA2008}&{GA}&CM& &$n\times k$&Criteria&{SA2008}&{GA}&CM \\
					
					\cline{1-5}\cline{7-11}
					\multirow{3}{*}{$16\times 6$} & ave$(|q|)$ &0.0098&0.0076&  & &\multirow{3}{*}{$18\times 8$}& ave$(|q|)$ &0.0241&0.0197& \\
					& max$|q|$ &0.0206&0.0118&No& & & max$|q|$   &0.0526&0.0361&No\\
					& CPU      &120&25&         & & & CPU        &195&42& \\
					
					\cline{1-5}\cline{7-11}
					\multirow{3}{*}{$17\times 6$} & ave$(|q|)$ &0.0101&0.0067&  & &\multirow{3}{*}{$22\times 8$}& ave$(|q|)$ &0.0211&0.0099& \\
					& max$|q|$ &0.0221&0.0147&Yes& & & max$|q|$   &0.0536&0.0220&No\\
					& CPU      &137&26&          & & & CPU        &231&43& \\
					
					\cline{1-5}\cline{7-11}
					\multirow{3}{*}{$18\times 6$} & ave$(|q|)$ &0.0118&0.0072&  & &\multirow{3}{*}{$28\times 8$}& ave$(|q|)$ &0.0155&0.0070& \\
					& max$|q|$ &0.0258&0.0114&No& & & max$|q|$   &0.0383&0.0142&No\\
					& CPU      &123&23&         & & & CPU        &274&40& \\
					
					\cline{1-5}\cline{7-11}
					\multirow{3}{*}{$17\times 8$} & ave$(|q|)$ &0.0207&0.0179&  & &\multirow{3}{*}{$30\times 8$}& ave$(|q|)$ &0.0150&0.0070& \\
					& max$|q|$ &0.0613&0.0343&Yes& & & max$|q|$   &0.0389&0.0131&No\\
					& CPU      &189&43&          & & & CPU        &293&42& \\
					\cline{1-5}\cline{7-11}
			\end{tabular}}
		\end{center}
	\end{table}

	\begin{table}\caption{Minimum ave$(|q|)$, max$|q|$, and Average CPU Time (in minutes) with Run Size Between $2^5$ and $2^8$.}\label{T10}
		\normalsize
		\begin{center}
			\begin{tabular}{|l|c|c|c|c|l|c|c|c|}
				\cline{1-4}\cline{6-9}
				$n\times k$&Criteria&{GA}&CM& &$n\times k$&Criteria&{GA}&CM \\
				
				\cline{1-4}\cline{6-9}       				
				\multirow{3}{*}{$33\times 8$}   & ave$(|q|)$ &0.0058&     &  &\multirow{3}{*}{$68\times 10$} & ave$(|q|)$ &0.0048&      \\
				& max$|q|$  &0.0124& Yes &  &   & max$|q|$   &0.0088& No   \\
				& CPU       &1.1   &     &  &   & CPU        &1.7   &      \\
				
				\cline{1-4}\cline{6-9}
				\multirow{3}{*}{$34\times 8$}   & ave$(|q|)$ &0.0048&     &  &\multirow{3}{*}{$96\times 12$} & ave$(|q|)$ &0.0096&      \\
				& max$|q|$  &0.0096 & No  &  &   & max$|q|$   &0.0178& No   \\
				& CPU       &1.1   &     &  &   & CPU        &1.8   &      \\
				
				\cline{1-4}\cline{6-9}
				\multirow{3}{*}{$48\times 10$}  & ave$(|q|)$ &0.0094&     &  &\multirow{3}{*}{$128\times 14$}& ave$(|q|)$ &0.0136&      \\
				& max$|q|$  &0.0156& No  &  &   & max$|q|$   &0.0241& No   \\
				& CPU       &1.7   &     &  &	& CPU        &3.0   &      \\
				
				\cline{1-4}\cline{6-9}
				\multirow{3}{*}{$64\times 10$}  & ave$(|q|)$ &0.0063&     &  &\multirow{3}{*}{$192\times 14$}& ave$(|q|)$ &0.0089&      \\
				& max$|q|$  &0.0114& No  &  &   & max$|q|$   &0.0170& No   \\
				& CPU       &1.7   &     &  &	& CPU        &3.9   &      \\
				
				\cline{1-4}\cline{6-9}
				\multirow{3}{*}{$65\times 10$}  & ave$(|q|)$ &0.0038&     &  &\multirow{3}{*}{$256\times 16$}& ave$(|q|)$ &0.0119&      \\
				& max$|q|$  &0.0087& Yes &  &   & max$|q|$   &0.0216& No   \\
				& CPU       &1.7   &     &  &   & CPU        &4.5   &      \\
				\cline{1-4}\cline{6-9}
			\end{tabular}
		\end{center}
	\end{table}

	\begin{table}\caption{Minimum ave$(|q|)$, max$|q|$, and Average CPU Time (in minutes) with the Rule of Thumb Sizes.}\label{T11}
		\begin{center}
			\resizebox{\textwidth}{!}{\begin{tabular}{|c|c|c|c|c|c|c|c|c|}
					\hline
					&$n \times k$&{20 $\times$ 2}&{30 $\times$ 3}&{40 $\times$ 4}&{50 $\times$ 5}&{60 $\times$ 6}&{70 $\times$ 7}&{80 $\times$ 8}\\
					\hline
					Criteria& &Min(CPU)&Min(CPU)&Min(CPU)&Min(CPU)&Min(CPU)&Min(CPU)&Min(CPU)\\
					\hline
					
					\multirow{5}{*}{ave$(|q|)$}
					& SA     &0(0.1)&0.0013(0.1)&0.0126(0.2)&0.0281(0.8)&0.0304(1.0)&0.0401(1.4)&0.0398(2.1)\\
					& SA2008 &0(1.1)&0.0002(2.0)&0.0008(3.8)&0.0024(10.0)&0.0054(14.4)&0.0066(20.7)&0.0091(30.3)\\
					& SLHD   &0(0.2)&0.0022(0.4)&0.0091(0.8)&0.0266(2.3)&0.0317(3.4)&0.0399(4.8)&0.0477(7.0)\\
					& LaPSO  &0(0.2)&0.0002(0.4)&0.0004(0.6)&0.0006(1.9)&0.0006(2.6)&0.0011(3.6)&0.0020(5.0)\\
					& GA     &0(0.1)&0.0002(0.2)&0.0004(0.2)&0.0004(0.5)&0.0006(0.6)&0.0011(0.8)&0.0011(1.1)\\
					
					\hline
					\multirow{5}{*}{max$|q|$}
					& SA     &0(0.1)&0.0020(0.1)&0.0218(0.2)&0.0514(0.8)&0.0702(1.0)&0.1045(1.4)&0.1000(2.1)\\
					& SA2008 &0(1.1)&0.0002(2.0)&0.0015(3.8)&0.0053(10.0)&0.0103(14.4)&0.0204(20.7)&0.0259(30.3)\\
					& SLHD   &0(0.2)&0.0038(0.4)&0.0165(0.8)&0.0477(2.3)&0.0704(3.4)&0.0891(4.8)&0.0996(7.0)\\
					& LaPSO  &0(0.2)&0.0002(0.4)&0.0008(0.6)&0.0009(1.9)&0.0013(2.6)&0.0022(3.6)&0.0040(5.0)\\
					& GA     &0(0.1)&0.0002(0.1)&0.0008(0.2)&0.0008(0.5)&0.0011(0.6)&0.0022(0.8)&0.0024(1.1)\\
					
					\hline
			\end{tabular}}
		\end{center}
	\end{table}

	\section{Conclusion and Recommendation}\label{Con}

    In this paper, we focus on constructing three commonly used types of optimal LHDs: maximin distance LHDs, maximum projection LHDs, and (nearly) orthogonal LHDs. We have summarized and compared several widely used search algorithms, including SA \citep{morris1995exploratory}, SA2008 \citep{joseph2008orthogonal}, SLHD \citep{ba2015optimal}, LaPSO \citep{chen2013optimizing}, and GA \citep{liefvendahl2006study}, along with various algebraic constructions \citep{wang2018optimal, ye1998orthogonal, cioppa2007efficient, sun2010construction, tang1993orthogonal, lin2009construction, butler2001optimal}. Our goal is to provide guidance for practitioners in selecting appropriate experimental designs. Algebraic constructions are preferred when available, particularly for large designs, whereas search algorithms are employed to generate near-optimal LHDs with flexible sizes.
	
	From the numerical studies presented in Section~\ref{Result}, GA emerges as the most reliable approach to generate near-optimal LHDs, offering a good balance between performance and computational time. GA often provides competitive or superior results for moderate and large design sizes. LaPSO also performs well, but it generally requires substantially more computational time. For constructing MaxPro LHDs,  the R package \R{MaxPro} \citep{MaxPro} is computationally fast, but its performance is often inferior to that of GA implemented in the R package \R{LHD} \citep{LHD}. We provide three tables summarizing the recommended methods for different selected design sizes in Tables~\ref{T12} to \ref{T14}. We note that the relative performance of these methods may depend on the design size, tuning parameters, and computational budget, and thus the recommendations should be interpreted as practical guidelines rather than definitive rankings.

    The search algorithms discussed in this paper can also be adapted to generate other types of efficient experimental designs. An interesting future research is to extend the scope to include additional design types, such as fractional factorial designs \citep{dean1999design}, supersaturated designs \citep{lin1993new}, and order-of-addition designs \citep{lin2019design}.
	
	\begin{table}\caption{Recommended Algorithm (or Method) for Maximin Distance LHDs under the $L_2$ Distance.}\label{T12}
		\begin{center}
			\resizebox{\textwidth}{!}{\begin{tabular}{|c|c|c|c|c|c|c|c|c|c|}
					\hline
					$n$&{$k=2$}&{$k=3$}&{$k=4$}&{$k=5$}&{$k=6$}&{$k=7$}&{$k=8$}&{$k=9$}&{$k=10$}\\
					\hline
					4 to 5&FastMm&GA&GA&&&&&&\\
					\hline
					6 to 7&GA&GA&GA&&&&&&\\
					\hline
					8 to 15&GA&GA&GA&GA&GA&&&&\\
					\hline
					16 to 31&GA&GA&GA&GA&GA&GA&GA&&\\
					\hline
					32 to 63&GA&GA&GA&GA&GA&GA&GA&GA&GA\\
					\hline
			\end{tabular}}
		\end{center}
	\end{table}

	\begin{table}\caption{Recommended Algorithm for Maximum Projection LHDs.}\label{T13}
		\begin{center}
			\resizebox{\textwidth}{!}{\begin{tabular}{|c|c|c|c|c|c|c|c|c|c|}
					\hline
					$n$&{$k=2$}&{$k=3$}&{$k=4$}&{$k=5$}&{$k=6$}&{$k=7$}&{$k=8$}&{$k=9$}&{$k=10$}\\
					\hline
					4 to 5&MaxPro&MaxPro&MaxPro&&&&&&\\
					\hline
					6&MaxPro&MaxPro&GA&&&&&&\\
					\hline
					7&MaxPro&GA&GA&&&&&&\\
					\hline
					8&MaxPro&GA&GA&GA&GA&&&&\\
					\hline
					9 to 15&GA&GA&GA&GA&GA&&&&\\
					\hline
					16 to 31&GA&GA&GA&GA&GA&GA&GA&&\\
					\hline
					32 to 63&GA&GA&GA&GA&GA&GA&GA&GA&GA\\
					\hline
			\end{tabular}}
		\end{center}
	\end{table}

	\begin{table}\caption{Recommended Algorithm (or Method) for Orthogonal and Nearly Orthogonal LHDs}\label{T14}
		\begin{center}
			\begin{tabular}{|c|c|c|c|}
				\hline
				$n$&{$k=2$}&{$k=3$}&{$k=4$}\\
				\hline
				4&Sun10&GA&GA\\
				\hline
				5&Sun10/Ye98&GA&GA\\
				\hline
				6&GA&GA&GA\\
				\hline
				7&GA&GA&SA2008\\
				\hline
				8&Sun10&GA&Sun10\\
				\hline
				9&Sun10&GA&Sun10/Ye98\\
				\hline
				10&GA&GA&GA\\
				\hline
				11&GA&SA2008&SA2008\\
				\hline
				12&Sun10&SA2008&SA2008\\
				\hline
				13&Sun10&GA&SA2008\\
				\hline
				14&GA&SA2008&GA\\
				\hline
				15&GA&GA&GA\\
				\hline
				16&Sun10&GA&Sun10\\
				\hline
				17&Sun10&GA&Sun10\\
				\hline
				18&GA&GA&GA\\
				\hline
				19&GA&GA&GA\\
				\hline
				$4r$ or $4r+1$&Sun10&&\\
				\hline
				$4r+2$ or $4r+3$&GA&&\\
				\hline
				
				$8r$ or $8r+1$&&&Sun10\\
				\hline
				
				\hline
			\end{tabular}
		\end{center}
	\end{table}
	
\section*{Acknowledgement}

This research was partially supported by the NSF Grant DMS-2311186.

\section*{Declarations}

\noindent{\bf Conflict of interest:} The authors have no conflicts of interest to report.

\bibliography{mybibfile}

@article{ye1998orthogonal,
	title={Orthogonal column {L}atin hypercubes and their application in computer experiments},
	author={Ye, K. Qian},
	journal={Journal of the American Statistical Association},
	volume={93},
	number={444},
	pages={1430--1439},
	year={1998},
	publisher={Taylor \& Francis}
}

@article{cioppa2007efficient,
	title={Efficient nearly orthogonal and space-filling {L}atin hypercubes},
	author={Cioppa, Thomas M and Lucas, Thomas W},
	journal={Technometrics},
	volume={49},
	number={1},
	pages={45--55},
	year={2007},
	publisher={Taylor \& Francis}
}

@article{tang1993orthogonal,
	title={Orthogonal array-based {L}atin hypercubes},
	author={Tang, Boxin},
	journal={Journal of the American statistical association},
	volume={88},
	number={424},
	pages={1392--1397},
	year={1993},
	publisher={Taylor \& Francis}
}

@article{lin2009construction,
	title={Construction of orthogonal and nearly orthogonal {L}atin hypercubes},
	author={Lin, C. Devon and Mukerjee, Rahul and Tang, Boxin},
	journal={Biometrika},
	volume={96},
	number={1},
	pages={243--247},
	year={2009},
	publisher={Oxford University Press}
}

@article{sun2010construction,
	title={Construction of orthogonal {L}atin hypercube designs with flexible run sizes},
	author={Sun, Fasheng and Liu, Min-Qian and Lin, Dennis K. J.},
	journal={Journal of Statistical Planning and Inference},
	volume={140},
	number={11},
	pages={3236--3242},
	year={2010},
	publisher={Elsevier}
}

@article{sun2009construction,
	title={Construction of orthogonal {L}atin hypercube designs},
	author={Sun, Fasheng and Liu, Min-Qian and Lin, Dennis K. J.},
	journal={Biometrika},
	volume={96},
	number={4},
	pages={971--974},
	year={2009},
	publisher={Oxford University Press}
}

@article{williams1949experimental,
	title={Experimental designs balanced for the estimation of residual effects of treatments},
	author={Williams, EJ},
	journal={Australian Journal of Chemistry},
	volume={2},
	number={2},
	pages={149--168},
	year={1949},
	publisher={CSIRO}
}

@article{butler2001optimal,
	title={Optimal and orthogonal {L}atin hypercube designs for computer experiments},
	author={Butler, Neil A},
	journal={Biometrika},
	volume={88},
	number={3},
	pages={847--857},
	year={2001},
	publisher={Oxford University Press}
}

@article{johnson1990minimax,
	title={Minimax and maximin distance designs},
	author={Johnson, Mark E and Moore, Leslie M and Ylvisaker, Donald},
	journal={Journal of statistical planning and inference},
	volume={26},
	number={2},
	pages={131--148},
	year={1990},
	publisher={Elsevier}
}

@article{morris1995exploratory,
	title={Exploratory designs for computational experiments},
	author={Morris, Max D and Mitchell, Toby J},
	journal={Journal of statistical planning and inference},
	volume={43},
	number={3},
	pages={381--402},
	year={1995},
	publisher={Elsevier}
}

@article{hickernell1998generalized,
	title={A generalized discrepancy and quadrature error bound},
	author={Hickernell, Fred},
	journal={Mathematics of computation},
	volume={67},
	number={221},
	pages={299--322},
	year={1998}
}

@article{joseph2008orthogonal,
	title={Orthogonal-maximin {L}atin hypercube designs},
	author={Joseph, V Roshan and Hung, Ying},
	journal={Statistica Sinica},
	pages={171--186},
	year={2008},
	publisher={JSTOR}
}

@article{jin2005efficient,
	title={An efficient algorithm for constructing optimal design of computer experiments},
	author={Jin, Ruichen and Chen, Wei and Sudjianto, Agus},
	journal={Journal of statistical planning and inference},
	volume={134},
	number={1},
	pages={268--287},
	year={2005},
	publisher={Elsevier}
}

@article{wang2018optimal,
	title={Optimal maximin ${L}_{1}$-distance Latin hypercube designs based on good lattice point designs},
	author={Wang, Lin and Xiao, Qian and Xu, Hongquan},
	journal={The Annals of Statistics},
	volume={46},
	number={6B},
	pages={3741--3766},
	year={2018},
	publisher={Institute of Mathematical Statistics}
}

@article{xiao2018construction,
	title={Construction of maximin distance designs via level permutation and expansion},
	author={Xiao, Qian and Xu, Hongquan},
	journal={Statistica Sinica},
	volume={28},
	number={3},
	pages={1395--1414},
	year={2018},
	publisher={JSTOR}
}

@article{xiao2017construction,
	title={Construction of maximin distance {L}atin squares and related {L}atin hypercube designs},
	author={Xiao, Qian and Xu, Hongquan},
	journal={Biometrika},
	volume={104},
	number={2},
	pages={455--464},
	year={2017},
	publisher={Oxford University Press}
}

@article{fang2002centered,
	title={Centered ${L}_{2}$-discrepancy of random sampling and {L}atin hypercube design, and construction of uniform designs},
	author={Fang, Kai-Tai and Ma, Chang-Xing and Winker, Peter},
	journal={Mathematics of Computation},
	volume={71},
	number={237},
	pages={275--296},
	year={2002}
}

@article{zhou2015space,
	title={Space-filling properties of good lattice point sets},
	author={Zhou, Yongdao and Xu, Hongquan},
	journal={Biometrika},
	volume={102},
	number={4},
	pages={959--966},
	year={2015},
	publisher={Oxford University Press}
}

@article{leary2003optimal,
	title={Optimal orthogonal-array-based {L}atin hypercubes},
	author={Leary, Stephen and Bhaskar, Atul and Keane, Andy},
	journal={Journal of Applied Statistics},
	volume={30},
	number={5},
	pages={585--598},
	year={2003},
	publisher={Taylor \& Francis}
}

@article{ba2015optimal,
	title={Optimal sliced {L}atin hypercube designs},
	author={Ba, Shan and Myers, William R and Brenneman, William A},
	journal={Technometrics},
	volume={57},
	number={4},
	pages={479--487},
	year={2015},
	publisher={Taylor \& Francis}
}

@article{qian2012sliced,
	title={Sliced {L}atin hypercube designs},
	author={Qian, Peter ZG},
	journal={Journal of the American Statistical Association},
	volume={107},
	number={497},
	pages={393--399},
	year={2012},
	publisher={Taylor \& Francis Group}
}

@article{chen2013optimizing,
	title={Optimizing {L}atin hypercube designs by particle swarm},
	author={Chen, Ray-Bing and Hsieh, Dai-Ni and Hung, Ying and Wang, Weichung},
	journal={Statistics and computing},
	volume={23},
	number={5},
	pages={663--676},
	year={2013},
	publisher={Springer}
}

@article{liefvendahl2006study,
	title={A study on algorithms for optimization of {L}atin hypercubes},
	author={Liefvendahl, Mattias and Stocki, Rafa{\l}},
	journal={Journal of statistical planning and inference},
	volume={136},
	number={9},
	pages={3231--3247},
	year={2006},
	publisher={Elsevier}
}

@article{joseph2015maximum,
	title={Maximum projection designs for computer experiments},
	author={Joseph, V Roshan and Gul, Evren and Ba, Shan},
	journal={Biometrika},
	volume={102},
	number={2},
	pages={371--380},
	year={2015},
	publisher={Oxford University Press}
}

@Manual{MaxPro,
	title = {MaxPro: Maximum Projection Designs},
	author = {Shan Ba and V. Roshan Joseph},
	year = {2018},
	note = {R package version 4.1-2},
	url = {https://CRAN.R-project.org/package=MaxPro},
}

@Manual{LHD,
	title = {LHD: Latin Hypercube Designs (LHDs)},
    author = {Hongzhi Wang and Qian Xiao and Abhyuday Mandal},
    year = {2025},
    note = {R package version 1.4.1},
    url = {https://CRAN.R-project.org/package=LHD},
}

@article{sacks1989designs,
	title={Designs for computer experiments},
	author={Sacks, Jerome and Schiller, Susannah B and Welch, William J},
	journal={Technometrics},
	volume={31},
	number={1},
	pages={41--47},
	year={1989},
	publisher={Taylor \& Francis Group}
}

@book{fang2005design,
	title={Design and modeling for computer experiments},
	author={Fang, Kai-Tai and Li, Runze and Sudjianto, Agus},
	year={2005},
	publisher={CRC press}
}

@article{mckay1979comparison,
	title={Comparison of three methods for selecting values of input variables in the analysis of output from a computer code},
	author={McKay, Michael D and Beckman, Richard J and Conover, William J},
	journal={Technometrics},
	volume={21},
	number={2},
	pages={239--245},
	year={1979},
	publisher={Taylor \& Francis}
}

@article{kenny2000algorithmic,
	title={Algorithmic construction of optimal symmetric {L}atin hypercube designs},
	author={Ye, K. Qian and Li, William and Sudjianto, Agus},
	journal={Journal of statistical planning and inference},
	volume={90},
	number={1},
	pages={145--159},
	year={2000},
	publisher={Elsevier}
}

@article{grosso2009finding,
	title={Finding maximin {L}atin hypercube designs by iterated local search heuristics},
	author={Grosso, Andrea and Jamali, ARMJU and Locatelli, Marco},
	journal={European Journal of Operational Research},
	volume={197},
	number={2},
	pages={541--547},
	year={2009},
	publisher={Elsevier}
}

@article{steinberg2006construction,
	title={A construction method for orthogonal {L}atin hypercube designs},
	author={Steinberg, David M and Lin, Dennis K. J.},
	journal={Biometrika},
	volume={93},
	number={2},
	pages={279--288},
	year={2006},
	publisher={Oxford University Press}
}

@article{yang2012construction,
	title={Construction of orthogonal and nearly orthogonal {L}atin hypercube designs from orthogonal designs},
	author={Yang, Jinyu and Liu, Min-Qian},
	journal={Statistica Sinica},
	pages={433--442},
	year={2012},
	publisher={JSTOR}
}

@article{georgiou2014some,
	title={Some classes of orthogonal {L}atin hypercube designs},
	author={Georgiou, Stelios D and Efthimiou, Ifigenia},
	journal={Statistica Sinica},
	volume={24},
	number={1},
	pages={101--120},
	year={2014},
	publisher={JSTOR}
}

@article{sun2017general,
	title={A general rotation method for orthogonal {L}atin hypercubes},
	author={Sun, Fasheng and Tang, Boxin},
	journal={Biometrika},
	volume={104},
	number={2},
	pages={465--472},
	year={2017},
	publisher={Oxford University Press}
}

@article{georgiou2009orthogonal,
	title={Orthogonal {L}atin hypercube designs from generalized orthogonal designs},
	author={Georgiou, Stelios D},
	journal={Journal of Statistical Planning and Inference},
	volume={139},
	number={4},
	pages={1530--1540},
	year={2009},
	publisher={Elsevier}
}

@inproceedings{kennedy1995particle,
	title={Particle swarm optimization},
	author={Kennedy, James and Eberhart, Russell},
	booktitle={Proceedings of ICNN'95-International Conference on Neural Networks},
	volume={4},
	pages={1942--1948},
	year={1995},
	organization={IEEE}
}

@book{holland1992adaptation,
	title={Adaptation in natural and artificial systems: an introductory analysis with applications to biology, control, and artificial intelligence},
	author={Holland, John Henry and others},
	year={1992},
	publisher={MIT press}
}

@book{goldberg1989genetic,
	author    = {Goldberg, David E.},
	title     = {Genetic Algorithms in Search, Optimization, and Machine Learning},
	publisher = {Addison-Wesley},
	year      = {1989}
}

@article{chen2015minimax,
	title={Minimax optimal designs via particle swarm optimization methods},
	author={Chen, Ray-Bing and Chang, Shin-Perng and Wang, Weichung and Tung, Heng-Chih and Wong, Weng Kee},
	journal={Statistics and Computing},
	volume={25},
	number={5},
	pages={975--988},
	year={2015},
	publisher={Springer}
}

@article{wong2015modified,
	title={A modified particle swarm optimization technique for finding optimal designs for mixture models},
	author={Wong, Weng Kee and Chen, Ray-Bing and Huang, Chien-Chih and Wang, Weichung},
	journal={PLoS One},
	volume={10},
	number={6},
	pages={e0124720},
	year={2015},
	publisher={Public Library of Science San Francisco, CA USA}
}

@article{kirkpatrick1983optimization,
	title={Optimization by simulated annealing},
	author={Kirkpatrick, Scott and Gelatt, C Daniel and Vecchi, Mario P},
	journal={science},
	volume={220},
	number={4598},
	pages={671--680},
	year={1983},
	publisher={American association for the advancement of science}
}

@article{loeppky2009choosing,
	title={Choosing the sample size of a computer experiment: A practical guide},
	author={Loeppky, Jason L and Sacks, Jerome and Welch, William J},
	journal={Technometrics},
	volume={51},
	number={4},
	pages={366--376},
	year={2009},
	publisher={Taylor \& Francis}
}

@article{chapman1994arctic,
	title={Arctic sea ice variability: Model sensitivities and a multidecadal simulation},
	author={Chapman, William L and Welch, William J and Bowman, Kenneth P and Sacks, Jerome and Walsh, John E},
	journal={Journal of Geophysical Research: Oceans},
	volume={99},
	number={C1},
	pages={919--935},
	year={1994},
	publisher={Wiley Online Library}
}

@article{jones1998efficient,
	title={Efficient global optimization of expensive black-box functions},
	author={Jones, Donald R and Schonlau, Matthias and Welch, William J},
	journal={Journal of Global optimization},
	volume={13},
	number={4},
	pages={455--492},
	year={1998},
	publisher={Springer}
}

@article{lin2019design,
	title={Design of order-of-addition experiments},
	author={Peng, Jiayu and Mukerjee, Rahul and Lin, Dennis K. J.},
	journal={Biometrika},
	volume={106},
	number={3},
	pages={683--694},
	year={2019},
	publisher={Oxford University Press}
}

@article{lin1993new,
	title={A new class of supersaturated designs},
	author={Lin, Dennis K. J.},
	journal={Technometrics},
	volume={35},
	number={1},
	pages={28--31},
	year={1993},
	publisher={Taylor \& Francis}
}

@book{dean1999design,
	title={Design and analysis of experiments},
	author={Dean, Angela and Voss, Daniel and Dragulji{\'c}, Danel and others},
	year={2017},
	publisher={Springer International Publishing}
}

@article{shewry1987maximum,
	title={Maximum entropy sampling},
	author={Shewry, Michael C and Wynn, Henry P},
	journal={Journal of applied statistics},
	volume={14},
	number={2},
	pages={165--170},
	year={1987},
	publisher={Taylor \& Francis}
}

@article{harari2018computer,
	title={Computer experiments: Prediction accuracy, sample size and model complexity revisited},
	author={Harari, Ofir and Bingham, Derek and Dean, Angela and Higdon, Dave},
	journal={Statistica Sinica},
	pages={899--919},
	year={2018},
	publisher={JSTOR}
}

@Manual{SLHD,
    title = {SLHD: Maximin-Distance (Sliced) Latin Hypercube Designs},
    author = {Shan Ba},
    year = {2015},
    note = {R package version 2.1-1},
    url = {https://CRAN.R-project.org/package=SLHD},
  }

@Manual{R,
    title = {R: A Language and Environment for Statistical Computing},
    author = {{R Core Team}},
    organization = {R Foundation for Statistical Computing},
    address = {Vienna, Austria},
    year = {2024},
    url = {https://www.R-project.org/},
  }

@article{wang2026rjournal,
  author = {Wang, Hongzhi and Xiao, Qian and Mandal, Abhyuday},
  title = {{LHD: An All-encompassing R Package for Constructing Optimal Latin Hypercube Designs}},
  journal = {The R Journal},
  year = {2026},
  doi = {10.32614/RJ-2025-033},
  volume = {17},
  issue = {4},
  issn = {2073-4859},
  pages = {20-36}
}
	
\end{document}